\documentclass[10pt,onecolumn]{IEEEtran}
\usepackage{amsthm}
\usepackage{amssymb}
\usepackage{graphicx}
\usepackage{multirow}
\usepackage[cmex10]{amsmath}
\usepackage{subfig}

\newcommand{\abs}[1]{\vert #1 \vert }

\newcommand{\defn}{\triangleq}

\newcommand{\msf}[1]{\mathsf{#1}}

\newtheorem{theorem}{Theorem}
\newtheorem{lemma}{Lemma}

\newtheorem{define}{Definition}

\newtheorem{proposition}{Proposition}
\allowdisplaybreaks
\begin{document}
\title{Detecting Byzantine Attacks for Gaussian Two-Way Relay System}

\author{Ruohan Cao
  \thanks{R.~Cao is with the
     the School of Information and Communication Engineering,
    Beijing University of Posts and Telecommunications (BUPT), Beijing
    100876, China (e-mail: caoruohan@bupt.edu.cn). }}
\maketitle

\begin{abstract}
This paper focuses on Byzantine attack detection for Gaussian two-way relay network. In this network, two source nodes communicate with each other with the help of an amplify-and-forward relay which may perform Byzantine attacks by forwarding altered symbols to the sources. For simple investigating the detectability of attacks conducted in Gaussian channels, we focus on the MA channel of the network, while assuming the BC channel is noiseless.
Upon such model, we propose a attack detection scheme implemented in the sources. Specifically, we consider a open wireless propagation environment that allows the symbols, forwarded by the relay, to go through a continuous channel and arrive to the sources. With the observations of the source, we develop a detection scheme for the source by comparing the joint empirical distribution of its received and transmitted signals with the known channel statistics. The main contribution of this paper is to prove that if and only if the Gaussian relay network satisfies a non-manipulable channel condition, the proposed detection scheme can detect arbitrary attacks that allows the stochastic distributions of altered symbols to vary arbitrarily and depend on each other. No pre-shared secret or secret transmission is needed for the detection. Furthermore, we also prove that for the considered Gaussian two-way relay networks, the non-manipulable channel condition is always satisfied. This result indicates that arbitrary attacks conducted in MA Gaussian channels are detectable by only using observations, while providing 
a base for attack detection in more general Gaussian networks. 
\end{abstract}
\section{Introduction}
Relay nodes are widely employed in modern communication networks to
  enhance coverage and connectivity of the networks. This dependence
  on the relaying infrastructure may increase the risk on security as
  malicious relays may forward false information in order to deceive the
  intended participants into accepting counterfeit information. These attacks, referred to as Byzantine
    attacks, impose significant ramifications on the design of
  network protocols \cite{Buttyan2006Security}\cite{bloch2011physical}. With the presence of Byzantine attacks, the attack detection technique, which 
determines whether Byzantine attacks are conducted or not, is one of the key steps 
supporting secure communication. 

 The work on attack detection starts above physical-layer, where
 each link is treated as a unit-capacity bit-pipe, while specific physical-layer characteristics are 
shielded. 
Based on this setting, 
cryptography keys are often used to make attacks detectable \cite{papadimitratos2006secure}, \cite{hu2005ariadne},
while requiring
    the cryptographic keys, to which the
    relays are not privy, to be shared between the source and
    destination before the communication takes place.
 Without using cryptography keys,
information theoretic detection schemes are proposed for multicast system or Caterpillar Network \cite{ho2008byzantine, kosut2009nonlinear}.  
These schemes are able to achieve errorless performance in probability, yet assuming that at least one relay or
link is absolutely trustworthy.


Besides these schemes treating channels as noiseless bit-pipes, there are
also many other attack detection schemes designed according to specific characteristics of physical-layer channel
for varying application scenarios. These schemes are mainly enabled by utilizing tracing symbols, or 
self-information provided by network topology structure.
In particular, source node inserts tracing symbols into a sequence of information bits, and then sends them together to
the destination.  After tracing symbols go through the relay channel, degraded by channel fading, noise, and
possible attacks, they are observed by the destination. 
Relying on the \emph{priori} knowledge of tracing symbols,
the destination could get the stochastic probability distributions of its observation on the tracing symbols, respectively under conditions
that the relay is reliable or malicious. On the basis of the two conditional statistical distributions, applying theory of hypothesis test,
attack detection schemes can be derived with perfect CSI \cite{mao2007tracing, Tradeoff} or no need of CSI \cite{noncoherent, nonherentsCL} for varying network scenarios .
The tracing-symbol based schemes commonly assume that the value and insertion location of the tracing symbols are known only at the source
and the destination, which indeed requires a additional  tracing-symbol distribution
mechanism implemented between communication parties. Besides that, since the explicit 
conditional statistical distributions are needed, these schemes assume that each malicious relay garbles
its received symbols according to independent and identically
distributed (i.i.d.) stochastic distributions. This model of
i.i.d. attacks may not always be valid in practice, although it makes
analysis simple. The Byzantine attack detection methods presented in
\cite{mao2007tracing}-\cite{nonherentsCL} may no longer be provably unbreakable for non-i.i.d. attacks.

Notice that all the above-mentioned schemes detect attacks by inserting redundancy, 
which increases the overhead cost.
In contrast, the schemes, which utilize side information (SI), do not need to insert any 
redundancy or just insert negligible redundancy. To be more specific, in \cite{OFDM}, for a wireless OFDMA
network, source node detects attacks from the correlation between its overheard signals from the relay 
and its transmitted signals. In \cite{KimTWC}, for the detection implemented in the destination of wireless one-way network, perfect error correction codes
(ECCs) are assumed to be used in the direct link between source and destination
and the relay link. With the help of using ECC in the direct link and relay link, the destination has opportunity 
to observe the exact information that are transmitted by the source or the relay. 
Then, the detection can be done by comparing the observations from the two links.
The detection performance of \cite{OFDM} and \cite{KimTWC} are impacted by channel fading, 
especially, the performance of \cite{KimTWC} depends
on the the quality of direct link, such that it may not work well in the network where direct link does not
exist or suffers deep fading.

On the contrary, \cite{he2013strong}-\cite{CaoTIFS} could probabilistically detect attacks. To be more specific, 
\cite{he2013strong}-\cite{GravesISIT13} consider two-hop communication for the typical three-node network composing
of an pair of communication parties and an signal one potential malicious relay, which is possible to conduct arbitrary attacks. 
To elaborate a little further, the scheme of \cite{he2013strong} requires the communication parties 
to transmit signal to relay simultaneously so as to support secret
transmission, with which some key components of algebraic manipulation detection (AMD) code
could be shared between communication parties, while keeping confidential to the relay.
Then, Byzantine attacks are detectable by applying AMD code to encrypt informations bits.
Even the AMD encryption cost negligible redundancy, the secret transmission increases the
overhead cost of the system.
This scheme is difficult to
extend this scheme to non-Gaussian channels. 
In our previous work
\cite{GravesINFOCOM12}-\cite{GravesISIT13}, focusing on two-way relay system, we show
that for discrete memoryless channels (DMCs), it is possible to detect potential Byzantine attacks
dispensing any AMD
code or cryptographic keys. The basic idea is 
that each node utilizes its own
transmitted symbols as clean reference for
statistically checking against the other node's symbols forwarded by the
relay. This scheme is difficult to
extend beyond DMC channels. 
We also extend the scheme of \cite{GravesINFOCOM12}-\cite{GravesISIT13} to the DMC two-hop relay system composing of a pair of communication parties and two potential malicious relays, where 
the all observations of the destination are prone to be attacked \cite{CaoTIFS}. No clean reference is available for the destination. 
This work relaxed the restriction imposed on the relay's misbehavior beyond i.i.d attacks for the DMC two-hop relay system. However, due to the lack of clean reference in DMC two-hop relay system, we cannot properly protect communication parties against arbitrary attacks. It indicates if all the signals observed by the destination are unsecured, the malicious relay have opportunity to fool the destination, which advocates the necessary of clean reference in attack detection.

This paper focuses on Byzantine attack detection for Gaussian two-way relay network. In two hops, two source nodes communicate with each other with the help of an amplify-and-forward relay which may perform Byzantine attacks by forwarding altered symbols to the sources. For simple investigating the detectability of attacks conducted in Gaussian channels, we focus on the MA channel of the network, while assuming the BC channel is noiseless.
The goal of this paper is to make communication parties probabilistically detect arbitrary attacks, despite of i.i.d or non-i.i.d attacks, without using any AMD code or secret transmission. 
In particular, notice that the source nodes' transmitted signals are not attacked and statistic related to its observation in the BC phase,  then we propose a detection method that use the 
transmitted signals as clean reference to statistically checking against
the signals observed from the relay. This model and method may seem a little bit similar to our previous work \cite{GravesINFOCOM12}-\cite{GravesISIT13}.
It is worth noting that in the Gaussian channel, the observations are continuous signals, the possible attacks are also continuous in the sense that they are likely to be conducted within continuous alphabet.
It is considerable difference with our previous work that are applicable for discrete alphabets \cite{GravesINFOCOM12}-\cite{GravesISIT13}. 
For the continuous attacks and channel model, the core contribution of this manuscript is to prove that the detectability of continuous attacks is equivalent to
non-manipulability of continuous channels. Also, we prove the considered Gaussian channel satisfies the non-manipulable condition. 
 This result indicates that arbitrary attacks conducted in MA Gaussian channels are detectable by only using observations, while providing 
a base for attack detection in more general Gaussian networks.

\section{system model}

Let us focus on the two-way relay example, where the two source nodes are time and phase synchronized. The two source nodes are termed as source 1 and source 2, respectively. The two sources exchange information with each other during two stages. Each stage includes $n$ instant. In the first $n$ instants, source 1 and source 2 respectively sends $n$-length sequences $X_1^{n}$ and $X_2^{n}$ to the relay node. $X_1$ and $X_2$ are equiprobability binary symbols generated from alphabet $(+1, -1)$. The MAC channel from the two sources to the relay is specified by $U = X_1 + X_2 + N_r$, where $U$ is the received signal of the relay in each instant, $N_r$ is AWGN existed in the MAC channel. Secondly, in the instants $n+1, n+2, \ldots, 2n$, the the relay forwarded $V^n$ to the two sources. Through a noiseless broadcast channel, source 1 receives sequence $Y_1^n=V^n$. $N_r$ is random variable following continuous stochastic distribution. Without loss of generalization, we assume the PDF of $N_r$ 
is $f_{N_r}\left(x\right)=\frac{1}{\sqrt{2\pi}}\exp\left(-{x^{2}}\right)$. Upon this assumption, the pdf of $U$ conditioned on $X_1=1$ and $X_1=-1$ can be given as $f_{U\left|-1\right.}\left(x\right)=\frac{0.5}{\sqrt{2\pi}}\exp\left(-{\left(x+2\right)^{2}}\right)+\frac{0.5}{\sqrt{2\pi}}\exp\left(-{x^{2}}\right)$ and $f_{U\left|+1\right.}\left(x\right)=\frac{0.5}{\sqrt{2\pi}}\exp\left(-{\left(x-2\right)^{2}}\right)+\frac{0.5}{\sqrt{2\pi}}\exp\left(-{x^{2}}\right)$. Then, the CDF of $U$ given $X_1=\mathsf{x}_{1}$ is 
$F_{U\left|\mathsf{x}_{1}\right.}\left(t\left|\mathsf{x}_{1}\right.\right)=\int_{-\infty}^{t}f_{U\left|\mathsf{x}_{1}\right.}\left(u\left|\mathsf{x}_{1}\right.\right)du$.

Let us assume $X_1^{n}$ and $X_2^{n}$ are i.i.d sequences and both MAC channel and BC channel are memoryless. Hence, $U^n$, $N_r^n$ are both i.i.d sequences. 
$V_n$ is the forwarded sequences. In this paper, the relay is allowed to conduct arbitrary attack, Then, the stochastic distribution of $V_n$ depends on the relay's behavior. In order to define the relay's behavior in mathematic sense,  let us choose one $n'$-length sequence $\mathsf{\widetilde{u}}_{1},\mathsf{\widetilde{u}}_{2},\ldots,\mathsf{\widetilde{u}}_{n'}$. Correspondingly, $\mathcal{B}\left(\mathsf{\widetilde{u}}_{j}\right)$ are domain consisting of $\mathsf{\widetilde{u}}_{j}$. They 
satisfy the constraints as follows.
\begin{align*}
&\alpha_{1}=\mathsf{\widetilde{u}}_{1}<\mathsf{\widetilde{u}}_{2}<\mathsf{\widetilde{u}}_{3}\ldots<\mathsf{\widetilde{u}}_{n'-1}=\beta_{1},\:\beta_{1}<\mathsf{\widetilde{u}}_{n'}\\
&\widetilde{\mathsf{u}}_{j}-\widetilde{\mathsf{u}}_{j-1}=\frac{\beta_{1}-\alpha_{1}}{n'-2},j=2,3,\ldots,n'-1\\
&\mathcal{B}\left(\widetilde{\mathsf{u}}_{j}\right)=\begin{cases}
\begin{array}{cc}
\left(\widetilde{\mathsf{u}}_{j-1},\,\widetilde{\mathsf{u}}_{j}\right], & j=2,3,\ldots,n'-1\\
\left(-\infty,\,\alpha_{1}\right], & j=1\\
\left(\beta_{1},\,+\infty\right), & j=n'
\end{array}\end{cases}
\end{align*}where $\alpha_{1}$ and $\beta_{1}$ are assumed to depend on $n'$. We will prove later that there exist a setup method to make $\alpha_{1}$, $\beta_{1}$ and $n'$ have the following properties.
\begin{equation}\label{constr1}
\lim_{n'\rightarrow\infty}F_{U\left|\mathsf{x}_{1}\right.}\left(\alpha_{1}\left|\mathsf{x}_{1}\right.\right)=0,\,\lim_{n'\rightarrow\infty}F_{U\left|\mathsf{x}_{1}\right.}\left(\beta_{1}\left|\mathsf{x}_{1}\right.\right)=1,\,\lim_{n'\rightarrow\infty}\frac{\beta_{1}-\alpha_{1}}{n'-2}=0.
\end{equation}

Based on the definition of $\mathcal{B}\left(\widetilde{\mathsf{u}}_{j}\right)$, the continuous variable $U$
can be quantized to discrete $\widetilde{U}$. In particular, if $U\in\mathcal{B}\left(\widetilde{\mathsf{u}}_{j}\right)$, then $\widetilde{U}=\widetilde{\mathsf{u}}_{j}$.
In other words, $\widetilde{U}\triangleq\sum_{j=1}^{n'}1\left(U\in\mathcal{B}\left(\widetilde{\mathsf{u}}_{j}\right)\right)\widetilde{\mathsf{u}}_{j}$.
For the random variable $\widetilde{U}$,  we use $\widetilde{\mathcal{U}}$ to denote its  alphabet. Also, we use $\widetilde{u}_i$ denote the generic symbol over $\widetilde{\mathcal{U}}$ in the $i$-th instant.
From these notations, the function $F^{(n')}_{V^{n}\left|\widetilde{U}^{n}\right.}\left(t\left|u\right.\right)$ for sequence pair ($U^n$, $V^n$) is defined as
\begin{equation}
F^{(n')}_{V^{n}\left|\widetilde{U}^{n}\right.}\left(t\left|u\right.\right)=\begin{cases}
\begin{array}{cc}
\frac{\sum_{i=1}^{n}1_{i}\left(v_{i}\leq t\right)1_{i}\left(\widetilde{u}_{i}=\widetilde{\mathsf{u}}_{j}\right)}{N\left(\widetilde{\mathsf{u}}_{j}\left|\widetilde{U}^{n}\right.\right)}, & N\left(\widetilde{\mathsf{u}}_{j}\left|\widetilde{U}^{n}\right.\right)\neq0,\, u\in\mathcal{B}\left(\widetilde{\mathsf{u}}_{j}\right)\\
0, & otherwise
\end{array}\end{cases}
\end{equation}
 By replacing $U$ and $V$ with their lower cases, the similar definition can be applied to 
$F^{(n')}_{v^{n}\left|\widetilde{u}^{n}\right.}\left(t\left|u\right.\right)$, which is definite function associated with particular sequence $(u^n,v^n)$. If the relay is absolutely reliable, we must always have {\small{$ \sum_{j=1}^{{n'-1}}\sum_{i=1}^{n'}|F_{V^{n}\left|\widetilde{U}^{n}\right.}^{\left(n'\right)}\left(\widetilde{\mathsf{u}}_{j}\left|\widetilde{\mathsf{u}}_{i}\right.\right)-\Phi\left(\widetilde{\mathsf{u}}_{j}-\widetilde{\mathsf{u}}_{i}\right)|=0$}}, where
 \begin{equation}
\Phi\left(t\right)=\begin{cases}
\begin{array}{cc}
1, & t \geq 0\\
0, & otherwise
\end{array}\end{cases}.
 \end{equation}
From this intuitively understanding, the malicious relay is defined
as follows. 
\begin{define} \label{def:maliciousness}
\textbf{(Malicious Relay)} The relay is said to be non-malicious if $\sum_{j=1}^{{n'-1}}\sum_{i=1}^{n'}|F_{V^{n}\left|\widetilde{U}^{n}\right.}^{\left(n'\right)}\left(\widetilde{\mathsf{u}}_{j}\left|\widetilde{\mathsf{u}}_{i}\right.\right)-\Phi\left(\widetilde{\mathsf{u}}_{j}-\widetilde{\mathsf{u}}_{i}\right)|
 \rightarrow 0$ in probability as $n$ and $n'$ approach to infinity. Otherwise,
  the relay is considered malicious.\end{define}
 Note that Definition~\ref{def:maliciousness} tolerates manipulating
only a negligible fraction of symbols by the relays. This relaxation
has essentially no effect on the information rate from the source to
the destination across the relays. Definition~\ref{def:maliciousness} also tolerates modification conducted by the relay 
within a negligible extent. This negligible extent is specified by $\frac{\beta_{1}-\alpha_{1}}{n'-2}$, as $\frac{\beta_{1}-\alpha_{1}}{n'-2}\rightarrow0$, 
this modification
has essentially no effect on the information reliability from the source to
the destination across the relay. 

For easy description of the main results on detectability, we give the expression of standard 
the empirical CDF of $v^{n}$ conditioned on $\mathsf{x_1}$ as
\begin{equation}
F_{v^{n}\left|x_1^{n}\right.}^{n}\left(t\left|\msf{x}_1\right.\right)=\frac{\sum_{i=1}^{n}1\left(v_{i}\leq t\right)1\left(x_{1,i}=\msf{x}_1\right)}{N(\msf{x}_1|x_1^{n})}, t\in\left(-\infty,+\infty\right).
 \end{equation}By replacing $v$ and $x_1$ with their upper cases,  random $F_{V^{n}\left|X_1^{n}\right.}^{n}\left(t\left|\msf{x}_1\right.\right)$ is similarly
 defined. 

\section{Main Results}
We first point that the wireless two-way relay channel is 
non-manipulable. 
  \begin{proposition}
  \textbf{Non-manipulability of Wireless Channel} The aforementioned channel is non-manipulable, which indicates there does not exist continuous functions $\Psi\left(v\left|u\right.\right)$ that satisfies the following three conditions
  \begin{enumerate}
  \item Fixing $u$, $\Psi\left(v\left|u\right.\right)$ is a pdf.
  \item For arbitrary value of $u$, $\int_{-\infty}^{+\infty}\left|\int_{-\infty}^{t}\Psi\left(v\left|u\right.\right)dv-\Phi\left(t-u\right)\right|^{2}dt$ has a strictly positive lower bound.
   \item {\small{$\int_{-\infty}^{+\infty}f_{U\left|\mathsf{x}_{1}\right.}\left(u\right)\Psi\left(v\left|u\right.\right)du=f_{U\left|\mathsf{x}_{1}\right.}\left(v\right)$}}.
  \end{enumerate}
\end{proposition}
\begin{theorem}\textbf{(Maliciousness detectability of wireless relay network)} \label{thm:main2}
 As a beneficial result of the fact that wireless MAC channel is non-manipulable,  there exist a
  sequence of decision statistics $\left\{ D^{n} \right\}$
  simultaneously having the following two properties: \\
  Fix any sufficiently small $\delta >0$, $\epsilon>0$, there has sufficiently large $n'$, 
\begin{enumerate}
\item {\small{$\lim_{n\rightarrow\infty}\Pr \left( D^{n} > \varepsilon(n',\delta)  ~\Big| 
   \sum_{j=1}^{{n'-1}}\sum_{i=1}^{n'}|F_{V^{n}\left|\widetilde{U}^{n}\right.}^{\left(n'\right)}\left(\widetilde{\mathsf{u}}_{j}\left|\widetilde{\mathsf{u}}_{i}\right.\right)-\Phi\left(\widetilde{\mathsf{u}}_{j}-\widetilde{\mathsf{u}}_{i}\right)|> \delta \right) \geq 1 -
  \epsilon$}} whenever {\small{$\Pr \Big(\sum_{j=1}^{{n'-1}}\sum_{i=1}^{n'}|F_{V^{n}\left|\widetilde{U}^{n}\right.}^{\left(n'\right)}\left(\widetilde{\mathsf{u}}_{j}\left|\widetilde{\mathsf{u}}_{i}\right.\right)-\Phi\left(\widetilde{\mathsf{u}}_{j}-\widetilde{\mathsf{u}}_{i}\right)|> \delta \Big) > 0$}}, where $\varepsilon(n',\delta)$ is strictly positive and can be arbitrary small. 
\item {\small{$\lim_{n\rightarrow\infty}\Pr\Big( D^{n}> \mu' (n',\delta) ~\Big| 
   \sum_{j=1}^{{n'-1}}\sum_{i=1}^{n'}|F_{V^{n}\left|\widetilde{U}^{n}\right.}^{\left(n'\right)}\left(\widetilde{\mathsf{u}}_{j}\left|\widetilde{\mathsf{u}}_{i}\right.\right)-\Phi\left(\widetilde{\mathsf{u}}_{j}-\widetilde{\mathsf{u}}_{i}\right)|\leq \delta \Big) \leq
  \epsilon$}} whenever {\small{$\Pr \Big( \sum_{j=1}^{{n'-1}}\sum_{i=1}^{n'}|F_{V^{n}\left|\widetilde{U}^{n}\right.}^{\left(n'\right)}\left(\widetilde{\mathsf{u}}_{j}\left|\widetilde{\mathsf{u}}_{i}\right.\right)-\Phi\left(\widetilde{\mathsf{u}}_{j}-\widetilde{\mathsf{u}}_{i}\right)|\leq \delta \Big) > 0$,
  where $\mu' (n',\delta) \rightarrow 0$ as $n'\rightarrow \infty$, $\delta \rightarrow 0$}}.
\end{enumerate}
\end{theorem}
\section{Proof of Proposition 1}
\begin{IEEEproof}
Let us assume the manipulable wireless channel exists, which indicates there at least one i.i.d attack making the statistical distribution of $U$ conditioned on $X_1$ is equivalent to the statistical distribution of $V$ conditioned on $X_1$. Hence, we have $I\left(X_{1};U\right)=I\left(X_{1};V\right)$, where $I\left(\cdot;\cdot\right)$ denotes mutual information between the two input variables. On the other hand, $\left(X_{1},U,V\right)$ forms a Markov chain as $X_{1}\rightarrow U\rightarrow V$. From Data-Processing Inequality,  $I\left(X_{1};U\right)=I\left(X_{1};V\right)$ implies the Markov chain $X_{1}\rightarrow V\rightarrow U$ is also established. Then, we have
\begin{equation}
\label{markov_eq}
\Pr\left(a<U<b\left|V=\mathsf{v},X_{1}=-1\right.\right)=\Pr\left(a<U<b\left|V=\mathsf{v},X_{1}=+1\right.\right)
 \end{equation}It is worth noting that due to the continuity of noise, for arbitrary value of $\mathsf{v}\in\left(-\infty,+\infty\right)$.
both $\Pr\left(a<U<b\left|V=\mathsf{v},X_{1}=-1\right.\right)$ and $\Pr\left(a<U<b\left|V=\mathsf{v},X_{1}=+1\right.\right)$ are well-defined. Furthermore, in this example $U = X_1 + X_2 + N$, {\small{$\Pr\left(a<U<b\left|V=\mathsf{v},X_{1}=-1\right.\right)$}} can be reshaped as 
\begin{equation}
\Pr\left(a<U<b\left|V=\mathsf{v},X_{1}=-1\right.\right)=\Pr\left(a+1<X_{2}+N<b+1\left|V=\mathsf{v},X_{1}=-1\right.\right).
\end{equation}Since $X_1$ is independent with $X_2$ and $N$, therefore we obtain
\begin{equation}
\Pr\left(a+1<X_{2}+N<b+1\left|V=\mathsf{v},X_{1}=-1\right.\right)=\Pr\left(a+1<X_{2}+N<b+1\left|V=\mathsf{v}\right.\right)
\end{equation}which indicates
\begin{equation}\label{reshape_1}
\Pr\left(a<U<b\left|V=\mathsf{v},X_{1}=-1\right.\right)=\Pr\left(a+1<X_{2}+N<b+1\left|V=\mathsf{v}\right.\right).
\end{equation}Similarly, we can have 
\begin{equation}\label{reshape_2}
\Pr\left(a<U<b\left|V=\mathsf{v},X_{1}=+1\right.\right)=\Pr\left(a-1<X_{2}+N<b-1\left|V=\mathsf{v}\right.\right).
\end{equation}Submitting (\ref{reshape_1}) and (\ref{reshape_2}) into (\ref{markov_eq}), we get
\begin{equation}
\Pr\left(a+1<X_{2}+N<b+1\left|V=\mathsf{v}\right.\right)=\Pr\left(a-1<X_{2}+N<b-1\left|V=\mathsf{v}\right.\right).
\end{equation}then, there has 
\begin{equation}\label{in_X2N}
\int_{-\infty}^{+\infty}\Pr\left(a+1<X_{2}+N<b+1\left|V=v\right.\right)f\left(v\right)dv=\int_{-\infty}^{+\infty}\Pr\left(a-1<X_{2}+N<b-1\left|V=v\right.\right)f\left(v\right)dv
\end{equation}It is not hard to find the two sides of (\ref{in_X2N}) are equivalent to $\Pr\left(a+1<X_{2}+N<b+1\right)$ and $\Pr\left(a-1<X_{2}+N<b-1\right)$, respectively. Finally, we have 
\begin{equation}\label{con}
\Pr\left(a+1<X_{2}+N<b+1\right)=\Pr\left(a-1<X_{2}+N<b-1\right).
\end{equation}
In summary,  under the assumption that the the channel is manipulable, the equation (\ref{con}) should be established for arbitrary $a$ and $b$. In other words, if we can find a pair $(a,b)$ for a wireless channel that $\Pr\left(a+1<X_{2}+N<b+1\right)\neq \Pr\left(a-1<X_{2}+N<b-1\right)$, then the
channel is non-manipulable. 
\end{IEEEproof}
Notice that the system model does not consider the noise of the BC channel.
Actually, even consider the noisy BC channel, the wireless network is till non-manipulable. This assertion is proved as follows. 
\begin{proposition}\label{lemp2}
For random variables $Z_1$, $Z_2$, $Z_3$ $Z_4$, $Z_5$, where
$Z_3=Z_1+Z_2$, $Z_5=Z_4+Z_2$, $Z_1$ and $Z_4$ are both stochastic independent 
with $Z_2$, 
if pdf $f_{Z_{3}\left|X_{1}\right.}\left({z}_{3}\left|\mathsf{x}_{1}\right.\right)=f_{Z_{5}\left|X_{1}\right.}\left({z}_{5}\left|\mathsf{x}_{1}\right.\right)$, then there must have $f_{Z_{1}\left|X_{1}\right.}\left({z}_{1}\left|\mathsf{x}_{1}\right.\right)=f_{Z_{4}\left|X_{1}\right.}\left({z}_{4}\left|\mathsf{x}_{1}\right.\right)$.
\end{proposition}
\begin{IEEEproof}
According to the fact that $Z_3=Z_1+Z_2$,  where $Z_1$ and $Z_2$ are stochastic independent 
with each other, then the characteristic function of $Z_3$ conditioned on $X_1=\mathsf{x}_{1}$ is expressed by 
{\small{\begin{equation}\label{c1}
\varphi_{Z_{3}\left|X_{1}\right.}\left(t\left|\mathsf{x}_{1}\right.\right)=\varphi_{Z_{1}\left|X_{1}\right.}\left(t\left|\mathsf{x}_{1}\right.\right)\varphi_{Z_{2}\left|X_{1}\right.}\left(t\left|\mathsf{x}_{1}\right.\right),
\end{equation}}}where $\varphi_{Z_{3}\left|X_{1}\right.}\left(t\left|\mathsf{x}_{1}\right.\right)$, $\varphi_{Z_{1}\left|X_{1}\right.}\left(t\left|\mathsf{x}_{1}\right.\right)$ and $\varphi_{Z_{2}\left|X_{1}\right.}\left(t\left|\mathsf{x}_{1}\right.\right)$ denote the characteristic functions of $Z_3$, $Z_2$ and $Z_1$ conditioned on $X_1=\mathsf{x}_{1}$, respectively.
Similarly, according to the fact that $Z_5=Z_4+Z_2$,  where $Z_4$ and $Z_2$ are stochastic independent 
with each other, then the characteristic function of $Z_5$ conditioned on $X_1=\mathsf{x}_{1}$ is expressed by
{\small{\begin{equation}\label{c2}
\varphi_{Z_{5}\left|X_{1}\right.}\left(t\left|\mathsf{x}_{1}\right.\right)=\varphi_{Z_{4}\left|X_{1}\right.}\left(t\left|\mathsf{x}_{1}\right.\right)\varphi_{Z_{2}\left|X_{1}\right.}\left(t\left|\mathsf{x}_{1}\right.\right),
\end{equation}}}where $\varphi_{Z_{5}\left|X_{1}\right.}\left(t\left|\mathsf{x}_{1}\right.\right)$ and $\varphi_{Z_{4}\left|X_{1}\right.}\left(t\left|\mathsf{x}_{1}\right.\right)$ denote the characteristic functions of $Z_4$ and $Z_2$ conditioned on $X_1=\mathsf{x}_{1}$, respectively.
Since {\small{$f_{Z_{3}\left|X_{1}\right.}\left(\mathsf{z}_{3}\left|\mathsf{x}_{1}\right.\right)=f_{Z_{5}\left|X_{1}\right.}\left(\mathsf{z}_{5}\left|\mathsf{x}_{1}\right.\right)$}}, we have 
{\small{\begin{equation}\label{c3}
\varphi_{Z_{3}\left|X_{1}\right.}\left(t\left|\mathsf{x}_{1}\right.\right)=\varphi_{Z_{5}\left|X_{1}\right.}\left(t\left|\mathsf{x}_{1}\right.\right)
\end{equation}}}Substituting (\ref{c1}) and (\ref{c2}) into (\ref{c1}), we get
{\small{\begin{equation}\label{c4}
\varphi_{Z_{4}\left|X_{1}\right.}\left(t\left|\mathsf{x}_{1}\right.\right)\varphi_{Z_{2}\left|X_{1}\right.}\left(t\left|\mathsf{x}_{1}\right.\right)=\varphi_{Z_{1}\left|X_{1}\right.}\left(t\left|\mathsf{x}_{1}\right.\right)\varphi_{Z_{2}\left|X_{1}\right.}\left(t\left|\mathsf{x}_{1}\right.\right).
\end{equation}}}Since {\small{$\varphi_{Z_{2}\left|X_{1}\right.}\left(t\left|\mathsf{x}_{1}\right.\right)$}}is characteristic function which always attains strictly non-zero value across $t\in\left(-\infty,+\infty\right)$, then we have 
{\small{\begin{equation}\label{c5}
\varphi_{Z_{4}\left|X_{1}\right.}\left(t\left|\mathsf{x}_{1}\right.\right)=\varphi_{Z_{1}\left|X_{1}\right.}\left(t\left|\mathsf{x}_{1}\right.\right).
\end{equation}}}From the knowledge that pdf can be uniquely determined by characteristic function, (\ref{c5}) indicates
{\small{\begin{equation}\label{c6}
f_{Z_{1}\left|X_{1}\right.}\left({z}_{1}\left|\mathsf{x}_{1}\right.\right)=f_{Z_{4}\left|X_{1}\right.}\left({z}_{4}\left|\mathsf{x}_{1}\right.\right).
\end{equation}}}
\end{IEEEproof}
Let us back to the proof of non-manipulability for the system with noisy BC channels. In such case, the source observes $Y =V+N_s$ in the BC phase, where $N_s$ denotes the noise of the BC channel.
Revisiting $Y =V+N_s$, based on proof firstly given in the section,
if and only if the relay is absolutely reliable, i.e., $U=V$, we can get {\small{$f_{U\left|X_{1}\right.}\left(u\left|\mathsf{x}_{1}\right.\right)=f_{V\left|X_{1}\right.}\left(v\left|\mathsf{x}_{1}\right.\right)$}}. 
Then, consider that Proposition \ref{lemp2} indicates if and only if {\small{$f_{U\left|X_{1}\right.}\left(u\left|\mathsf{x}_{1}\right.\right)=f_{V\left|X_{1}\right.}\left(v\left|\mathsf{x}_{1}\right.\right)$}}, there exists {\small{$f_{U+N_s\left|X_{1}\right.}\left(y\left|\mathsf{x}_{1}\right.\right)=f_{V+N_s\left|X_{1}\right.}\left(y\left|\mathsf{x}_{1}\right.\right)$}}. We finally get that if and only if $U=V$, {\small{$f_{U+N_s\left|X_{1}\right.}\left(y\left|\mathsf{x}_{1}\right.\right)=f_{V+N_s\left|X_{1}\right.}\left(y\left|\mathsf{x}_{1}\right.\right)$}} holds true. Hence, we have proved the non-manipulability of the wireless network having noisy BC channel.
The above-mentioned proof indicates our work could be applicable for the wireless network where MA and BC channels are both noisy.
For simplicity, we give the detailed proof for the wireless network where only MA channel is noisy.

\section{Proof of Theorem~\ref{thm:main2}: Preparations}
In order to prove one convergence property of $F^{(n')}_{V^{n}\left|\widetilde{U}^{n}\right.}\left(t\left|u\right.\right)$ given later, we also
define 
{\small{\begin{align}
&\nonumber\triangle F_{i,i',j,j'}=P_{X_{i},X_{i'}\left|V_{i},V_{i'},\widetilde{U}_{i},\widetilde{U}_{i'}\right.}\left\{ \mathsf{x}_{1},\,\mathsf{x}_{1}\left|v_{i}\leq t,\, v_{i'}\leq t',\,\widetilde{\mathsf{u}}_{j},\,\widetilde{\mathsf{u}}_{j'}\right.\right\} -P_{X_{1}\left|\widetilde{U}\right.}\left(\mathsf{x}_{1}\left|\widetilde{\mathsf{u}}_{j}\right.\right)P_{X_{i'}\left|V_{i},V_{i'},\widetilde{U}_{i},\widetilde{U}_{i'}\right.}\left\{ \mathsf{x}_{1}\left|v_{i}\leq t,\, v_{i'}\leq t,\,\widetilde{\mathsf{u}}_{j},\,\widetilde{\mathsf{u}}_{j'}\right.\right\}\\\nonumber
&-P_{X_{i}\left|V_{i},V_{i'},\widetilde{U}_{i},\widetilde{U}_{i'}\right.}\left\{ \mathsf{x}_{1}\left|v_{i}\leq t,\, v_{i'}\leq t,\,\widetilde{\mathsf{u}}_{j},\,\widetilde{\mathsf{u}}_{j'}\right.\right\} P_{X_{1}\left|\widetilde{U}\right.}\left(\mathsf{x}_{1}\left|\widetilde{\mathsf{u}}_{j'}\right.\right)+P_{X_{1}\left|\widetilde{U}\right.}\left(\mathsf{x}_{1}\left|\widetilde{\mathsf{u}}_{j}\right.\right)P_{X_{1}\left|\widetilde{U}\right.}\left(\mathsf{x}_{1}\left|\widetilde{\mathsf{u}}_{j'}\right.\right)
\end{align}}}where $i\neq i'$, $j,j'=1,2,\ldots n'$, $i,i'=1,2,\ldots n$. Then, we have the following lemma.
\begin{lemma}\label{lem1}
If we choose $\alpha_{1}=-\beta_{1}$, and 
\begin{equation}\label{number}
\frac{1}{n'-2}=\max \{ \underbrace{P_{X_{1}\left|U\right.}\left(1\left|-\beta_{1}\right.\right)}_{\xi_{1}\left(\beta_{1}\right)},\underbrace{1-P_{X_{1}\left|U\right.}\left(1\left|\beta_{1}\right.\right)}_{\xi_{2}\left(\beta_{1}\right)},\underbrace{P_{X_{1}\left|U\right.}\left(-1\left|\beta_{1}\right.\right)}_{\xi_{3}\left(\beta_{1}\right)},\underbrace{1-P_{X_{1}\left|U\right.}\left(-1\left|-\beta_{1}\right.\right)}_{\xi_{4}\left(\beta_{1}\right)}\}. 
\end{equation}then upon this setup, besides the statement of (\ref{constr1}) is ture, 
there also exist a upper bound for $F_{i,i'j,j'}$ across $j,j'=1,2,\ldots n'$, $i,i'=1,2,\ldots n$, $i\neq i'$. This upper bound only depends on $n'$ rather than $n$. Hence, we denote the upper bound as $\triangle F_{max}\left(n'\right)$. $\triangle F_{max}\left(n'\right)$ has property that  
\begin{equation}
\lim_{n'\rightarrow\infty}\left(\beta_{1}-\alpha_{1}\right)^{k}n'F_{max}\left(n'\right)\rightarrow0
\end{equation}where $k$ is a bounded integer.
\end{lemma}
\begin{IEEEproof}
To that end, we first reshape $F_{i,i',j,j'}$ as 
{\small{\begin{align}
&\nonumber\triangle F_{i,i',j,j'}=P_{X_{1,i'}\left|V_{i},V_{i'},\widetilde{U}_{i},\widetilde{U}_{i'}\right.}\left\{ \mathsf{x}_{1}\left|v_{i}\leq t,\, v_{i'}\leq t,\,\widetilde{\mathsf{u}}_{j},\,\widetilde{\mathsf{u}}_{j'}\right.\right\} \left(P_{X_{1,i}\left|V_{i},V_{i'},\widetilde{U}_{i},\widetilde{U}_{i'}\right.}\left\{ \mathsf{x}_{1}\left|v_{i}\leq t,\, v_{i'}\leq t,\,\widetilde{\mathsf{u}}_{j},\,\widetilde{\mathsf{u}}_{j'}\right.\right\} -P_{X_{1}\left|\widetilde{U}\right.}\left(\mathsf{x}_{1}\left|\widetilde{\mathsf{u}}_{j}\right.\right)\right)\\
&\nonumber-P_{X_{1}\left|\widetilde{U}\right.}\left(\mathsf{x}_{1}\left|\widetilde{\mathsf{u}}_{j'}\right.\right)\left(P_{X_{1,i}\left|V_{i},V_{i'},\widetilde{U}_{i},\widetilde{U}_{i'}\right.}\left\{ \mathsf{x}_{1}\left|v_{i}\leq t,\, v_{i'}\leq t,\,\widetilde{\mathsf{u}}_{j},\,\widetilde{\mathsf{u}}_{j'}\right.\right\} -P_{X_{1}\left|\widetilde{U}\right.}\left(\mathsf{x}_{1}\left|\widetilde{\mathsf{u}}_{j}\right.\right)\right)\\
&=\left(P_{X_{1,i}\left|V_{i},V_{i'},\widetilde{U}_{i},\widetilde{U}_{i'}\right.}\left\{ \mathsf{x}_{1}\left|v_{i}\leq t,\, v_{i'}\leq t,\,\widetilde{\mathsf{u}}_{j},\,\widetilde{\mathsf{u}}_{j'}\right.\right\} -P_{X_{1}\left|\widetilde{U}\right.}\left(\mathsf{x}_{1}\left|\widetilde{\mathsf{u}}_{j}\right.\right)\right)\left(P_{X_{1,i'}\left|V_{i},V_{i'},\widetilde{U}_{i},\widetilde{U}_{i'}\right.}\left\{ \mathsf{x}_{1}\left|v_{i}\leq t,\, v_{i'}\leq t,\,\widetilde{\mathsf{u}}_{j},\,\widetilde{\mathsf{u}}_{j'}\right.\right\} -P_{X_{1}\left|\widetilde{U}\right.}\left(\mathsf{x}_{1}\left|\widetilde{\mathsf{u}}_{j'}\right.\right)\right),\label{reshape}
\end{align}}}where the first equality follows $X_1^n$ is i.i.d sequence, $X_{1,i'}$ is independent on $X_{1,i}$ due to $i\neq i'$. To bound $\triangle F_{i,i'j,j'}$, we have
\begin{equation}
P_{X_{1}\left|\widetilde{U}\right.}\left(\mathsf{x}_{1}\left|\widetilde{\mathsf{u}}_{j}\right.\right)=\frac{\int_{u\in\mathcal{B}\left(\widetilde{\mathsf{u}}_{j}\right)}P_{X_{1}\left|U\right.}\left(\mathsf{x}_{1}\left|u\right.\right)f_{U}\left(u\right)du}{\int_{u\in\mathcal{B}\left(\widetilde{\mathsf{u}}_{j}\right)}f_{U}\left(u\right)du}
\end{equation}which indicates
\begin{equation}\label{bound1}
\underset{u\in\mathcal{B}\left(\widetilde{\mathsf{u}}_{j}\right)}{\min}P_{X_{1}\left|U\right.}\left(\mathsf{x}_{1}\left|u\right.\right)\leq P_{X_{1}\left|\widetilde{U}\right.}\left(\mathsf{x}_{1}\left|\widetilde{\mathsf{u}}_{j}\right.\right)\leq\underset{u\in\mathcal{B}\left(\widetilde{\mathsf{u}}_{j}\right)}{\max}P_{X_{1}\left|U\right.}\left(\mathsf{x}_{1}\left|u\right.\right).
\end{equation}On the other hand, since
{\small{\begin{align*}
&P_{X_{1,i}\left|V_{i},V_{i'},\widetilde{U}_{i},\widetilde{U}_{i'}\right.}\left\{ \mathsf{x}_{1}\left|v_{i}\leq t,\, v_{i'}\leq t,\,\widetilde{\mathsf{u}}_{j},\,\widetilde{\mathsf{u}}_{j'}\right.\right\} =\\
&\frac{\int_{-\infty}^{t}\int_{-\infty}^{t}\int_{u\in\mathcal{B}\left(\widetilde{\mathsf{u}}_{j}\right)}\int_{u'\in\mathcal{B}\left(\widetilde{\mathsf{u}}_{j'}\right)}P_{X_{1}\left|U\right.}\left(\mathsf{x}_{1}\left|u\right.\right)f_{U_{i},U_{i'}\left|V_{i},V_{i'}\right.}\left(u,u'\left|v,\, v'\right.\right)f_{V_{i},V_{i'}}\left(v_{i},\, v_{i'}\right)dudu'dvdv'}{\int_{-\infty}^{t}\int_{-\infty}^{t}\int_{u\in\mathcal{B}\left(\widetilde{\mathsf{u}}_{j}\right)}\int_{u'\in\mathcal{B}\left(\widetilde{\mathsf{u}}_{j'}\right)}f_{U_{i},U_{i'}\left|V_{i},V_{i'}\right.}\left(u,u'\left|v,\, v'\right.\right)f_{V_{i},V_{i'}}\left(v_{i},\, v_{i'}\right)dudu'dvdv'}
\end{align*}}}we have 
\begin{equation}\label{bound2}
\underset{u\in\mathcal{B}\left(\widetilde{\mathsf{u}}_{j}\right)}{\min}P_{X_{1}\left|U\right.}\left(\mathsf{x}_{1}\left|u\right.\right)\leq P_{X_{i}\left|V_{i},V_{i'},\widetilde{U}_{i},\widetilde{U}_{i'}\right.}\left\{ \mathsf{x}_{1}\left|v_{i}\leq t,\, v_{i'}\leq t,\,\widetilde{\mathsf{u}}_{j},\,\widetilde{\mathsf{u}}_{j'}\right.\right\} \leq\underset{u\in\mathcal{B}\left(\widetilde{\mathsf{u}}_{j}\right)}{\max}P_{X_{1}\left|U\right.}\left(\mathsf{x}_{1}\left|u\right.\right)
\end{equation}Jointly considering (\ref{bound1}) and (\ref{bound2}), we have
\begin{equation}\label{bound3}
\left|P_{X_{1}\left|\widetilde{U}\right.}\left(\mathsf{x}_{1}\left|\widetilde{\mathsf{u}}_{j}\right.\right)-P_{X_{1,i}\left|V_{i},V_{i'},\widetilde{U}_{i},\widetilde{U}_{i'}\right.}\left\{ \mathsf{x}_{1}\left|v_{i}\leq t,\, v_{i'}\leq t,\,\widetilde{\mathsf{u}}_{j},\,\widetilde{\mathsf{u}}_{j'}\right.\right\} \right|<\underset{u\in\mathcal{B}\left(\widetilde{\mathsf{u}}_{j}\right)}{\max}P_{X_{1}\left|U\right.}\left(\mathsf{x}_{1}\left|u\right.\right)-\underset{u\in\mathcal{B}\left(\widetilde{\mathsf{u}}_{j}\right)}{\min}P_{X_{1}\left|U\right.}\left(\mathsf{x}_{1}\left|u\right.\right)
\end{equation}Following the similar logic from (\ref{bound1}) to (\ref{bound3}), we also have 
{\small{\begin{equation}\label{bound31}
\left|P_{X_{1,i'}\left|V_{i},V_{i'},\widetilde{U}_{i},\widetilde{U}_{i'}\right.}\left\{ \mathsf{x}_{1}\left|v_{i}\leq t,\, v_{i'}\leq t,\,\widetilde{\mathsf{u}}_{j},\,\widetilde{\mathsf{u}}_{j'}\right.\right\} -P_{X_{1}\left|\widetilde{U}\right.}\left(\mathsf{x}_{1}\left|\widetilde{\mathsf{u}}_{j'}\right.\right)\right|\leq\underset{u\in\mathcal{B}\left(\widetilde{\mathsf{u}}_{j'}\right)}{\max}P_{X_{1}\left|U\right.}\left(\mathsf{x}_{1}\left|u\right.\right)-\underset{u\in\mathcal{B}\left(\widetilde{\mathsf{u}}_{j'}\right)}{\min}P_{X_{1}\left|U\right.}\left(\mathsf{x}_{1}\left|u\right.\right)
\end{equation}}}Substituting (\ref{bound3}) and (\ref{bound31}) into (\ref{reshape}), we have
\begin{equation}\label{bound4}
\abs{\triangle F_{i,i',j,j'}}<\left(\underset{u\in\mathcal{B}\left(\widetilde{\mathsf{u}}_{j}\right)}{\max}P_{X_{1}\left|U\right.}\left(\mathsf{x}_{1}\left|u\right.\right)-\underset{u\in\mathcal{B}\left(\widetilde{\mathsf{u}}_{j}\right)}{\min}P_{X_{1}\left|U\right.}\left(\mathsf{x}_{1}\left|u\right.\right)\right)\left(\underset{u\in\mathcal{B}\left(\widetilde{\mathsf{u}}_{j'}\right)}{\max}P_{X_{1}\left|U\right.}\left(\mathsf{x}_{1}\left|u\right.\right)-\underset{u\in\mathcal{B}\left(\widetilde{\mathsf{u}}_{j'}\right)}{\min}P_{X_{1}\left|U\right.}\left(\mathsf{x}_{1}\left|u\right.\right)\right), 
\end{equation}and 
\begin{equation}\label{Fmax}
\underset{j,j'=1,2,\ldots n',i,i'=1,2,\ldots n,i\neq i'}{\max}\triangle F_{i,i'j,j'}<\underset{j=1,2,\ldots n'}{\max}\left(\underset{u\in\mathcal{B}\left(\widetilde{\mathsf{u}}_{j}\right)}{\max}P_{X_{1}\left|U\right.}\left(\mathsf{x}_{1}\left|u\right.\right)-\underset{u\in\mathcal{B}\left(\widetilde{\mathsf{u}}_{j}\right)}{\min}P_{X_{1}\left|U\right.}\left(\mathsf{x}_{1}\left|u\right.\right)\right)^2\defn\triangle F_{max}\left(n'\right)
\end{equation}Then, we proceed to focus on the property of $F_{max}\left(n'\right)$. 
Revisiting the system model, we have
{\small{\begin{equation}
P_{X_{1}\left|U\right.}\left(1\left|u\right.\right)=\frac{1}{2+\exp\left(4u-4\right)+\exp\left(-4u-4\right)}+\frac{1}{2\exp\left(4-4u\right)+1+\exp\left(-8u\right)}
\end{equation}}}and
{\small{\begin{equation}
P_{X_{1}\left|U\right.}\left(-1\left|u\right.\right)=\frac{1}{2+\exp\left(4u-4\right)+\exp\left(-4u-4\right)}+\frac{1}{2\exp\left(4+4u\right)+1+\exp\left(8u\right)}
\end{equation}}}According to $P_{X_{1}\left|U\right.}\left(1\left|u\right.\right)$, $P_{X_{1}\left|U\right.}\left(-1\left|u\right.\right)$ and (\ref{number}), 
we get $\lim_{n'\rightarrow\infty}\frac{\beta_{1}-\alpha_{1}}{n'-2}=0$. The statement of (\ref{constr1}) is thus proved. 
Furthermore,  in (\ref{number}) we have $\lim_{\beta_{1}\rightarrow\infty}\left(2\beta_{1}\right)^{k+2}\xi_{i}\left(\beta_{1}\right)=0$ and $\lim_{\beta_{1}\rightarrow\infty}\left(2\beta_{1}\right)^{k}\xi_{i}\left(\beta_{1}\right)=0$ for $i=1,2,3,4.$.
With these properties on limitation, 
for $j=2,3,\ldots,n'-1$, we have
{\small{\begin{equation}\
\left(\underset{u\in\mathcal{B}\left(\widetilde{\mathsf{u}}_{j}\right)}{\max}P_{X_{1}\left|U\right.}\left(\mathsf{x}_{1}\left|u\right.\right)-\underset{u\in\mathcal{B}\left(\widetilde{\mathsf{u}}_{j}\right)}{\min}P_{X_{1}\left|U\right.}\left(\mathsf{x}_{1}\left|u\right.\right)\right)^{2}\leq {P'}^{2}_{X_{1}\left|U\right.}\left(\mathsf{x}_{1}\left|u'_{j}\right.\right)\left(\frac{2\beta_{1}}{n'-2}\right)^{2}
\end{equation}}}where $u'_{j}\in\mathcal{B}\left(\widetilde{\mathsf{u}}_{j}\right)$, ${P'}_{X_{1}\left|U\right.}\left(\mathsf{x}_{1}\left|u\right.\right)$ is derived function of ${P}_{X_{1}\left|U\right.}\left(\mathsf{x}_{1}\left|u\right.\right)$. The maximum of ${P'}_{X_{1}\left|U\right.}\left(\mathsf{x}_{1}\left|u\right.\right)$ in $\left(-\infty,+\infty\right)$ is bounded. 
\begin{align}\label{limitation1}
&\nonumber\lim_{n'\rightarrow\infty}\left(2\beta_{1}\right)^{k}n'\left(\underset{u\in\mathcal{B}\left(\widetilde{\mathsf{u}}_{j}\right)}{\max}P_{X_{1}\left|U\right.}\left(\mathsf{x}_{1}\left|u\right.\right)-\underset{u\in\mathcal{B}\left(\widetilde{\mathsf{u}}_{j}\right)}{\min}P_{X_{1}\left|U\right.}\left(\mathsf{x}_{1}\left|u\right.\right)\right)^{2}\leq\lim_{n'\rightarrow\infty}\left(2\beta_{1}\right)^{k}n'{P'}^2_{X_{1}\left|U\right.}\left(\mathsf{x}_{1}\left|u'_{j}\right.\right)\left(\frac{2\beta_{1}}{n'-2}\right)^{2}\\&
=\lim_{\beta_{1}\rightarrow\infty}{P'}^2_{X_{1}\left|U\right.}\left(\mathsf{x}_{1}\left|u'_{j}\right.\right)\frac{\left(2\beta_{1}\right)^{k+2}}{n'-2}=0
\end{align}where the last equality follows the fact that the maximum of ${P'}_{X_{1}\left|U\right.}\left(\mathsf{x}_{1}\left|u\right.\right)$ in $\left(-\infty,+\infty\right)$ is bounded and in (\ref{number}) we have $\lim_{\beta_{1}\rightarrow\infty}\left(2\beta_{1}\right)^{k+2}\xi_{i}\left(\beta_{1}\right)=0$ for $i=1,2,3,4.$
Then, for $j=1$ and $j=n'$,  
\begin{equation}
\left(\underset{u\in\mathcal{B}\left(\widetilde{\mathsf{u}}_{j}\right)}{\max}P_{X_{1}\left|U\right.}\left(\mathsf{x}_{1}\left|u\right.\right)-\underset{u\in\mathcal{B}\left(\widetilde{\mathsf{u}}_{j}\right)}{\min}P_{X_{1}\left|U\right.}\left(\mathsf{x}_{1}\left|u\right.\right)\right)^{2}\leq\left(\frac{1}{n'-2}\right)^{2}
\end{equation}Hence, for $j=1$ and $j=n'$, we have
{\small{\begin{equation}\label{limitation2}
\lim_{n'\rightarrow\infty}\left(2\beta_{1}\right)^{k}n'\left(\underset{u\in\mathcal{B}\left(\widetilde{\mathsf{u}}_{j}\right)}{\max}P_{X_{1}\left|U\right.}\left(\mathsf{x}_{1}\left|u\right.\right)-\underset{u\in\mathcal{B}\left(\widetilde{\mathsf{u}}_{j}\right)}{\min}P_{X_{1}\left|U\right.}\left(\mathsf{x}_{1}\left|u\right.\right)\right)^{2}\leq\lim_{n'\rightarrow\infty}\left(2\beta_{1}\right)^{k}n'\left(\frac{1}{n'-2}\right)^{2}=\lim_{\beta_{1}\rightarrow\infty}\frac{\left(2\beta_{1}\right)^{k}}{n'-2}=0
\end{equation}}}where the last equality follows in (\ref{number}) we have $\lim_{\beta_{1}\rightarrow\infty}\left(2\beta_{1}\right)^{k}\xi_{i}\left(\beta_{1}\right)=0$ for $i=1,2,3,4.$ 
Combining (\ref{limitation1}) and (\ref{limitation2}), we get for each $j=1,2,\ldots,n'$, 
{\small{\begin{equation}
\lim_{n'\rightarrow\infty}\left(2\beta_{1}\right)^{k}n'\left(\underset{u\in\mathcal{B}\left(\widetilde{\mathsf{u}}_{j}\right)}{\max}P_{X_{1}\left|U\right.}\left(\mathsf{x}_{1}\left|u\right.\right)-\underset{u\in\mathcal{B}\left(\widetilde{\mathsf{u}}_{j}\right)}{\min}P_{X_{1}\left|U\right.}\left(\mathsf{x}_{1}\left|u\right.\right)\right)^{2}=0
\end{equation}}}Finally, based on the definition of $F_{max}\left(n'\right)$ in (\ref{Fmax}), the statement of this lemma is immediate. 
\end{IEEEproof}
Upon this lemma, we have the following convergence property.
\begin{lemma}\label{lem2}
For arbitrary $t$, sufficiently small $\mu$ and $\varepsilon\leq\frac{\mu}{2n'}$, 
{\small{\begin{align}
&\nonumber \Pr\left\{ \left|F_{V^{n}\left|X_{1}^{n}\right.}^{n}\left(t\left|\msf{x}_{1}\right.\right)-\int_{-\infty}^{+\infty}f_{U\left|X_{1}\right.}\left(u\left|\msf{x}_{1}\right.\right)F^{(n')}_{V^{n}\left|\widetilde{U}^{n}\right.}\left(t\left|u\right.\right)du\right|>\mu\right\} \\\label{final_upper}
&<\frac{4}{\mu^{2}}\left(\frac{n'^{2}}{n}+\frac{1}{\Pr\left\{ \left(X_{1}^{n},\,\widetilde{U}^{n}\right)\in\mathcal{A}_{\varepsilon}\right\} \left(P_{X_{1}}\left(\mathsf{x}\right)-\varepsilon\right)^{2}}\triangle F_{max}\left(n'\right)\right)+\Pr\left\{ \left(X_{1}^{n},\,\widetilde{U}^{n}\right)\notin T_{\left[X_{1},\widetilde{U}\right]_{\varepsilon}}^{n}\right\}
\end{align}}}
where{\small{\begin{equation}
\mathcal{A}_{\varepsilon}=\left\{ \left(x_{1}^{n},\widetilde{u}^{n}\right):\:\left|P_{\widetilde{U}\left|X_{1}\right.}\left(\widetilde{\mathsf{u}}\left|\mathsf{x}_{1}\right.\right)-\frac{N\left(\widetilde{\mathsf{u}}\left|\widetilde{U}^{n}\right.\right)P_{X_{1}\left|\widetilde{U}\right.}\left(\mathsf{x}_{1}\left|\widetilde{\mathsf{u}}\right.\right)}{N\left(\mathsf{x}_{1}\left|X_{1}^{n}\right.\right)}\right|<\varepsilon,\;\left|P_{X_{1}}\left(\mathsf{x}_{1}\right)-\frac{N\left(\mathsf{x}_{1}\left|X_{1}^{n}\right.\right)}{n}\right|<\varepsilon\right\}.
\end{equation}}}
\end{lemma}
\begin{IEEEproof}
Notice that
{\small{\begin{align}\label{upper}
&\Pr\left\{ \left|F_{V^{n}\left|X_{1}^{n}\right.}^{n}\left(t\left|\msf{x}_{1}\right.\right)-\int_{-\infty}^{+\infty}f_{U\left|X_{1}\right.}\left(u\left|\msf{x}_{1}\right.\right)F^{(n')}_{V^{n}\left|\widetilde{U}^{n}\right.}\left(t\left|u\right.\right)du\right|>\mu\right\} \\\nonumber
&<\Pr\left\{ \left|F_{V^{n}\left|X_{1}^{n}\right.}^{n}\left(t\left|\msf{x}_{1}\right.\right)-\int_{-\infty}^{+\infty}f_{U\left|X_{1}\right.}\left(u\left|\msf{x}_{1}\right.\right)F^{(n')}_{V^{n}\left|\widetilde{U}^{n}\right.}\left(t\left|u\right.\right)du\right|>\mu \left| \left(X_{1}^{n},\,\widetilde{U}^{n}\right)\in \mathcal{A}_{\varepsilon}\right.\right\} +\Pr\left\{ \left(X_{1}^{n},\,\widetilde{U}^{n}\right)\notin \mathcal{A}_{\varepsilon}\right\} 
\end{align}}}The proof of this lemma is equivalent to prove the two items in the right side of (\ref{upper}) both approach to $0$.
Firstly notice that after $n'$, $\alpha_{1}$ and $\beta_{1}$ are chosen and fixed properly,
\begin{equation}
\Pr\left\{ \left(X_{1}^{n},\,\widetilde{U}^{n}\right)\notin \mathcal{A}_{\varepsilon}\right\} \rightarrow0
\end{equation}as $n$ approaches to infinity.  Then, focusing on the first item in the right side of (\ref{upper}), $\left(X_{1}^{n},\,\widetilde{U}^{n}\right)\in \mathcal{A}_{\varepsilon}$ indicates $N\left(\widetilde{\mathsf{u}}_{j}\left|\widetilde{U}^{n}\right.\right)>0$ for all $j=1,2,\ldots, n'$. 
Hence, $\frac{\sum_{i=1}^{n}1_{i}\left(v_{i}\leq t\right)1_{i}\left(\widetilde{u}_{i}=\widetilde{\mathsf{u}}_{j}\right)}{N\left(\widetilde{\mathsf{u}}_{j}\left|\widetilde{U}^{n}\right.\right)}$ is well-defined for all $j=1,2,\ldots, n'$. Under the condition {\small{$ \left(X_{1}^{n},\,\widetilde{U}^{n}\right)\in \mathcal{A}_{\varepsilon}$}}, we have
{\small{\begin{align}\label{upper1}
&\left|F_{V^{n}\left|X_{1}^{n}\right.}^{n}\left(t\left|\msf{x}_{1}\right.\right)-\int_{-\infty}^{+\infty}f_{U\left|X_{1}\right.}\left(u\left|\msf{x}_{1}\right.\right)F^{(n')}_{V^{n}\left|\widetilde{U}^{n}\right.}\left(t\left|u\right.\right)du\right|\\\nonumber
&=\left|\frac{\sum_{i=1}^{n}1_{i}\left(v_{i}\leq t\right)1_{i}\left(x_{i}=\mathsf{x}_{1}\right)}{N\left(\mathsf{x}_{1}\left|X_{1}^{n}\right.\right)}-\sum_{j=1}^{n'}\frac{\sum_{i=1}^{n}1_{i}\left(v_{i}\leq t\right)1_{i}\left(\widetilde{u}_{i}=\widetilde{\mathsf{u}}_{j}\right)}{N\left(\widetilde{\mathsf{u}}_{j}\left|\widetilde{U}^{n}\right.\right)}\int_{\mathcal{B}\left(\widetilde{\mathsf{u}}_{j}\right)}f_{U\left|X_{1}\right.}\left(u\left|\msf{x}_{1}\right.\right)du\right|\\\nonumber
&\text{=}\left|\frac{\sum_{j=1}^{n'}\sum_{i=1}^{n}1_{i}\left(v_{i}\leq t\right)1_{i}\left(x_{i}=\mathsf{x}_{1}\right)1_{i}\left(\widetilde{u}_{i}=\widetilde{\mathsf{u}}_{j}\right)}{N\left(\mathsf{x}_{1}\left|X_{1}^{n}\right.\right)}-\sum_{j=1}^{n'}\frac{\sum_{i=1}^{n}1_{i}\left(v_{i}\leq t\right)1_{i}\left(\widetilde{u}_{i}=\widetilde{\mathsf{u}}_{j}\right)}{N\left(\widetilde{\mathsf{u}}_{j}\left|\widetilde{U}^{n}\right.\right)}P_{\widetilde{U}\left|X_{1}\right.}\left(\widetilde{\mathsf{u}}_{j}\left|\mathsf{x}_{1}\right.\right)\right|
\end{align}}}Substituting (\ref{upper1}) into the first item in the right side of (\ref{upper}), it becomes
{\small{\begin{align}\label{upper2}
&\Pr\left\{ \left|\frac{\sum_{j=1}^{n'}\sum_{i=1}^{n}1_{i}\left(v_{i}\leq t\right)1_{i}\left(x_{i}=\mathsf{x}_{1}\right)1_{i}\left(\widetilde{u}_{i}=\widetilde{\mathsf{u}}_{j}\right)}{N\left(\mathsf{x}_{1}\left|X_{1}^{n}\right.\right)}-\sum_{j=1}^{n'}\frac{\sum_{i=1}^{n}1_{i}\left(v_{i}\leq t\right)1_{i}\left(\widetilde{u}_{i}=\widetilde{\mathsf{u}}_{j}\right)}{N\left(\widetilde{\mathsf{u}}_{j}\left|\widetilde{U}^{n}\right.\right)}P_{\widetilde{U}\left|X_{1}\right.}\left(\widetilde{\mathsf{u}}_{j}\left|\mathsf{x}_{1}\right.\right)\right|>\mu\left|\left(X_{1}^{n},\,\widetilde{U}^{n}\right)\in\mathcal{A}_{\varepsilon}\right.\right\} \\\nonumber
&<\Pr\left\{ \left|\sum_{j=1}^{n'}\underbrace{\left(\frac{\sum_{i=1}^{n}1_{i}\left(v_{i}\leq t\right)1_{i}\left(x_{i}=\mathsf{x}_{1}\right)1_{i}\left(\widetilde{u}_{i}=\widetilde{\mathsf{u}}_{j}\right)}{n\left(P_{X_{1}}\left(\mathsf{x}\right)-\varepsilon\right)}-\frac{\sum_{i=1}^{n}1_{i}\left(v_{i}\leq t\right)1_{i}\left(\widetilde{u}_{i}=\widetilde{\mathsf{u}}_{j}\right)P_{X_{1}\left|\widetilde{U}\right.}\left(\mathsf{x}_{1}\left|\widetilde{\mathsf{u}}_{j}\right.\right)}{n\left(P_{X_{1}}\left(\mathsf{x}\right)-\varepsilon\right)}\right)}_{H_{j}}\right|>\frac{\mu}{2}\left|\left(X_{1}^{n},\,\widetilde{U}^{n}\right)\in\mathcal{A}_{\varepsilon}\right.\right\}  \\\nonumber
&+\sum_{j=1}^{n'}\Pr\left\{ \left|\frac{\sum_{i=1}^{n}1_{i}\left(v_{i}\leq t\right)1_{i}\left(\widetilde{u}_{i}=\widetilde{\mathsf{u}}_{j}\right)}{N\left(\widetilde{\mathsf{u}}_{j}\left|\widetilde{U}^{n}\right.\right)}P_{\widetilde{U}\left|X_{1}\right.}\left(\widetilde{\mathsf{u}}_{j}\left|\mathsf{x}_{1}\right.\right)-\frac{\sum_{i=1}^{n}1_{i}\left(v_{i}\leq t\right)1_{i}\left(\widetilde{u}_{i}=\widetilde{\mathsf{u}}_{j}\right)P_{X_{1}\left|\widetilde{U}\right.}\left(\mathsf{x}_{1}\left|\widetilde{\mathsf{u}}_{j}\right.\right)}{N\left(\mathsf{x}_{1}\left|X_{1}^{n}\right.\right)}\right|>\frac{\mu}{2n'}\left|\left(X_{1}^{n},\,\widetilde{U}^{n}\right)\in\mathcal{A}_{\varepsilon}\right.\right\} 
\end{align}}}The second item in the right of (\ref{upper2}) can be further bound as 
{\small{\begin{align}\label{upper3}
&\Pr\left\{ \left|\frac{\sum_{i=1}^{n}1_{i}\left(v_{i}\leq t\right)1_{i}\left(\widetilde{u}_{i}=\widetilde{\mathsf{u}}_{j}\right)}{N\left(\widetilde{\mathsf{u}}_{j}\left|\widetilde{U}^{n}\right.\right)}P_{\widetilde{U}\left|X_{1}\right.}\left(\widetilde{\mathsf{u}}_{j}\left|\mathsf{x}_{1}\right.\right)-\frac{\sum_{i=1}^{n}1_{i}\left(v_{i}\leq t\right)1_{i}\left(\widetilde{u}_{i}=\widetilde{\mathsf{u}}_{j}\right)P_{X_{1}\left|\widetilde{U}\right.}\left(\mathsf{x}_{1}\left|\widetilde{\mathsf{u}}_{j}\right.\right)}{N\left(\mathsf{x}_{1}\left|X_{1}^{n}\right.\right)}\right|>\frac{\mu}{2n'}\left|\left(X_{1}^{n},\,\widetilde{U}^{n}\right)\in \mathcal{A}_{\varepsilon}\right.\right\} \\\nonumber
&\leq\Pr\left\{ \left|P_{\widetilde{U}\left|X_{1}\right.}\left(\widetilde{\mathsf{u}}_{j}\left|\mathsf{x}_{1}\right.\right)-\frac{N\left(\widetilde{\mathsf{u}}\left|\widetilde{U}^{n}\right.\right)P_{X_{1}\left|\widetilde{U}\right.}\left(\mathsf{x}_{1}\left|\widetilde{\mathsf{u}}_{j}\right.\right)}{N\left(\mathsf{x}_{1}\left|X_{1}^{n}\right.\right)}\right|>\frac{\mu}{2n'}\left|\left(X_{1}^{n},\,\widetilde{U}^{n}\right)\in \mathcal{A}_{\varepsilon}\right.\right\} =0
\end{align}}}where the last equality follows the definition of $\mathcal{A}_{\varepsilon}$ and set of $\varepsilon\leq\frac{\mu}{2n'}$. From (\ref{upper3}), the second item in the right of (\ref{upper2}) equals to 0. Then, we proceed to bound the first item in the right of (\ref{upper2}) as
{\small{\begin{equation}\label{upper4}
\Pr\left\{ \left|\sum_{j=1}^{n'}H_{j}\right|>\frac{\mu}{2}\left|\left(X_{1}^{n},\,\widetilde{U}^{n}\right)\in\mathcal{A}_{\varepsilon}\right.\right\} <\frac{4}{\mu^{2}}E_{\mathcal{A}_{\varepsilon}}\left|\sum_{j=1}^{n'}H_{j}\right|^{2} =\frac{4}{\mu^{2}}\sum_{j=1}^{n'}\sum_{j'=1}^{n'}E_{\mathcal{A}_{\varepsilon}}\left(H_{j}H_{j'}\right)
\end{equation}}}which follows the Chebyshev theorem. $E_{\mathcal{A}_{\varepsilon}}\left(\cdot\right)$ indicates the expectation of its input conditioned on $\left(X_{1}^{n},\,\widetilde{U}^{n}\right)\in\mathcal{A}_{\varepsilon}$.
{\small{\begin{align}
&\nonumber E_{\mathcal{A}_{\varepsilon}}\left(H_{j}H_{j'}\right)=\\\nonumber
&\frac{E_{\mathcal{A}_{\varepsilon}}\left(\sum_{i=1}^{n}1_{i}\left(v_{i}\leq t\right)1_{i}\left(\widetilde{u}_{i}=\widetilde{\mathsf{u}}_{j}\right)\left(1_{i}\left(x_{1,i}=\mathsf{x}_{1}\right)-P_{X_{1}\left|\widetilde{U}\right.}\left(\mathsf{x}_{1}\left|\widetilde{\mathsf{u}}_{j}\right.\right)\right)\right)\left(\sum_{i=1}^{n}1_{i}\left(v_{i}\leq t\right)1_{i}\left(\widetilde{u}_{i}=\widetilde{\mathsf{u}}_{j'}\right)\left(1_{i}\left(x_{1,i}=\mathsf{x}_{1}\right)-P_{X_{1}\left|\widetilde{U}\right.}\left(\mathsf{x}_{1}\left|\widetilde{\mathsf{u}}_{j'}\right.\right)\right)\right)}{n^{2}\left(P_{X_{1}}\left(\mathsf{x}\right)-\varepsilon\right)^{2}}\\
&\nonumber \leq\frac{\sum_{i=1}^{n}E_{\mathcal{A}_{\varepsilon}}\left\{ 1_{i}\left(v_{i}\leq t\right)1_{i}\left(\widetilde{u}_{i}=\widetilde{\mathsf{u}}_{j}\right)\left(1_{i}\left(x_{1,i}=\mathsf{x}_{1}\right)-P_{X_{1}\left|\widetilde{U}\right.}\left(\mathsf{x}_{1}\left|\widetilde{\mathsf{u}}_{j}\right.\right)\right)1_{i}\left(\widetilde{u}_{i}=\widetilde{\mathsf{u}}_{j'}\right)\left(1_{i}\left(x_{1,i}=\mathsf{x}_{1}\right)-P_{X_{1}\left|\widetilde{U}\right.}\left(\mathsf{x}_{1}\left|\widetilde{\mathsf{u}}_{j'}\right.\right)\right)\right\} }{n^{2}\left(P_{X_{1}}\left(\mathsf{x}\right)-\varepsilon\right)^{2}}\\
&\nonumber +\frac{E\sum_{i=1}^{n}\sum_{i=1,i'\neq i}^{n}1_{i}\left(v_{i}\leq t\right)1_{i}\left(\widetilde{u}_{i}=\widetilde{\mathsf{u}}_{j}\right)1_{i'}\left(v_{i'}\leq t\right)1_{i'}\left(\widetilde{u}_{i'}=\widetilde{\mathsf{u}}_{j'}\right)\left(1_{i}\left(x_{1,i}=\mathsf{x}_{1}\right)-P_{X_{1}\left|\widetilde{U}\right.}\left(\mathsf{x}_{1}\left|\widetilde{\mathsf{u}}_{j}\right.\right)\right)\left(1_{i'}\left(x_{1,i'}=\mathsf{x}_{1}\right)-P_{X_{1}\left|\widetilde{U}\right.}\left(\mathsf{x}_{1}\left|\widetilde{\mathsf{u}}_{j'}\right.\right)\right)}{\Pr\left\{ \left(X_{1}^{n},\,\widetilde{U}^{n}\right)\in\mathcal{A}_{\varepsilon}\right\} n^{2}\left(P_{X_{1}}\left(\mathsf{x}\right)-\varepsilon\right)^{2}}\\\nonumber
&\leq\frac{1}{n}+\frac{\sum_{i=1}^{n}\sum_{i=1,i'\neq i}^{n}E\left\{ 1_{i}\left(v_{i}\leq t\right)1_{i}\left(\widetilde{u}_{i}=\widetilde{\mathsf{u}}_{j}\right)1_{i'}\left(v_{i'}\leq t\right)1_{i'}\left(\widetilde{u}_{i'}=\widetilde{\mathsf{u}}_{j'}\right)1_{i}\left(x_{1,i}=\mathsf{x}_{1}\right)1_{i'}\left(x_{1,i'}=\mathsf{x}_{1}\right)\right\} }{\Pr\left\{ \left(X_{1}^{n},\,\widetilde{U}^{n}\right)\in\mathcal{A}_{\varepsilon}\right\} n^{2}\left(P_{X_{1}}\left(\mathsf{x}\right)-\varepsilon\right)^{2}}\\\nonumber
&-\frac{\sum_{i=1}^{n}\sum_{i=1,i'\neq i}^{n}E\left\{ 1_{i}\left(v_{i}\leq t\right)1_{i}\left(\widetilde{u}_{i}=\widetilde{\mathsf{u}}_{j}\right)1_{i'}\left(v_{i'}\leq t\right)1_{i'}\left(\widetilde{u}_{i'}=\widetilde{\mathsf{u}}_{j'}\right)1_{i'}\left(x_{1,i'}=\mathsf{x}_{1}\right)\right\} P_{X_{1}\left|\widetilde{U}\right.}\left(\mathsf{x}_{1}\left|\widetilde{\mathsf{u}}_{j}\right.\right)}{\Pr\left\{ \left(X_{1}^{n},\,\widetilde{U}^{n}\right)\in\mathcal{A}_{\varepsilon}\right\} n^{2}\left(P_{X_{1}}\left(\mathsf{x}\right)-\varepsilon\right)^{2}}\\\nonumber
&-\frac{\sum_{i=1}^{n}\sum_{i=1,i'\neq i}^{n}E\left\{ 1_{i}\left(v_{i}\leq t\right)1_{i}\left(\widetilde{u}_{i}=\widetilde{\mathsf{u}}_{j}\right)1_{i'}\left(v_{i'}\leq t\right)1_{i'}\left(\widetilde{u}_{i'}=\widetilde{\mathsf{u}}_{j'}\right)1_{i}\left(x_{1,i}=\mathsf{x}_{1}\right)\right\} P_{X_{1}\left|\widetilde{U}\right.}\left(\mathsf{x}_{1}\left|\widetilde{\mathsf{u}}_{j'}\right.\right)}{\Pr\left\{ \left(X_{1}^{n},\,\widetilde{U}^{n}\right)\in\mathcal{A}_{\varepsilon}\right\} n^{2}\left(P_{X_{1}}\left(\mathsf{x}\right)-\varepsilon\right)^{2}}\\\nonumber
&+\frac{\sum_{i=1}^{n}\sum_{i=1,i'\neq i}^{n}E\left\{ 1_{i}\left(v_{i}\leq t\right)1_{i}\left(\widetilde{u}_{i}=\widetilde{\mathsf{u}}_{j}\right)1_{i'}\left(v_{i'}\leq t\right)1_{i'}\left(\widetilde{u}_{i'}=\widetilde{\mathsf{u}}_{j'}\right)\right\} P_{X_{1}\left|\widetilde{U}\right.}\left(\mathsf{x}_{1}\left|\widetilde{\mathsf{u}}_{j}\right.\right)P_{X_{1}\left|\widetilde{U}\right.}\left(\mathsf{x}_{1}\left|\widetilde{\mathsf{u}}_{j'}\right.\right)}{\Pr\left\{ \left(X_{1}^{n},\,\widetilde{U}^{n}\right)\in\mathcal{A}_{\varepsilon}\right\} n^{2}\left(P_{X_{1}}\left(\mathsf{x}\right)-\varepsilon\right)^{2}}\\\label{upper5}
&=\frac{1}{n}+\frac{\sum_{i=1}^{n}\sum_{i=1,i'\neq i}^{n}P_{V_{i},V_{i'},\widetilde{U}_{i},\widetilde{U}_{i'}}\left\{ v_{i}\leq t,\, v_{i'}\leq t,\,\widetilde{\mathsf{u}}_{j},\,\widetilde{\mathsf{u}}_{j}\right\} \triangle F_{i,i'j,j'}}{\Pr\left\{ \left(X_{1}^{n},\,\widetilde{U}^{n}\right)\in\mathcal{A}_{\varepsilon}\right\} n^{2}\left(P_{X_{1}}\left(\mathsf{x}\right)-\varepsilon\right)^{2}}
\end{align}}}Substituting (\ref{upper5}) into (\ref{upper4}), we have
{\small{\begin{align}
&\nonumber\Pr\left\{ \left|\sum_{j=1}^{n'}H_{j}\right|>\frac{\mu}{2}\left|\left(X_{1}^{n},\,\widetilde{U}^{n}\right)\in\mathcal{A}_{\varepsilon}\right.\right\} <\frac{4}{\mu^{2}}\sum_{j=1}^{n'}\sum_{j'=1}^{n'}\left\{ \frac{1}{n}+\frac{\sum_{i=1}^{n}\sum_{i=1,i'\neq i}^{n}P_{V_{i},V_{i'},\widetilde{U}_{i},\widetilde{U}_{i'}}\left\{ v_{i}\leq t,\, v_{i'}\leq t,\,\widetilde{\mathsf{u}}_{j},\,\widetilde{\mathsf{u}}_{j}\right\} }{\Pr\left\{ \left(X_{1}^{n},\,\widetilde{U}^{n}\right)\in\mathcal{A}_{\varepsilon}\right\} n^{2}\left(P_{X_{1}}\left(\mathsf{x}\right)-\varepsilon\right)^{2}}\triangle F_{max}\left(n'\right)\right\} \\\nonumber
&\leq\frac{4}{\mu^{2}}\left(\frac{n'^{2}}{n}+\frac{\sum_{i=1}^{n}\sum_{i=1,i'\neq i}^{n}\sum_{j=1}^{n'}\sum_{j'=1}^{n'}P_{\widetilde{U}}\left\{ \widetilde{\mathsf{u}}_{j}\right\} P_{\widetilde{U}}\left\{ \widetilde{\mathsf{u}}_{j'}\right\} }{\Pr\left\{ \left(X_{1}^{n},\,\widetilde{U}^{n}\right)\in\mathcal{A}_{\varepsilon}\right\} n^{2}\left(P_{X_{1}}\left(\mathsf{x}\right)-\varepsilon\right)^{2}}\triangle F_{max}\left(n'\right)\right)\\\label{upper6}
&\leq \frac{4}{\mu^{2}}\left(\frac{n'^{2}}{n}+\frac{1}{\Pr\left\{ \left(X_{1}^{n},\,\widetilde{U}^{n}\right)\in\mathcal{A}_{\varepsilon}\right\} \left(P_{X_{1}}\left(\mathsf{x}\right)-\varepsilon\right)^{2}}\triangle F_{max}\left(n'\right)\right)
\end{align}}}From (\ref{upper6}) (\ref{upper3}) (\ref{upper2}) and (\ref{upper}), we have 
\begin{align}
&\nonumber \Pr\left\{ \left|F_{V^{n}\left|X_{1}^{n}\right.}^{n}\left(t\left|\msf{x}_{1}\right.\right)-\int_{-\infty}^{+\infty}f_{U\left|X_{1}\right.}\left(u\left|\msf{x}_{1}\right.\right)F^{(n')}_{V^{n}\left|\widetilde{U}^{n}\right.}\left(t\left|u\right.\right)du\right|>\mu\right\} \\\label{final_upper}
&<\frac{4}{\mu^{2}}\left(\frac{n'^{2}}{n}+\frac{1}{\Pr\left\{ \left(X_{1}^{n},\,\widetilde{U}^{n}\right)\in\mathcal{A}_{\varepsilon}\right\} \left(P_{X_{1}}\left(\mathsf{x}\right)-\varepsilon\right)^{2}}\triangle F_{max}\left(n'\right)\right)+\Pr\left\{ \left(X_{1}^{n},\,\widetilde{U}^{n}\right)\notin T_{\left[X_{1},\widetilde{U}\right]_{\varepsilon}}^{n}\right\}
\end{align}The proof is finished.
\end{IEEEproof}
Upon the aforementioned lemmas, the following assertion is immediate. 
\begin{lemma}\label{lem3}
For sequence $t_{1}, t_{2}, \ldots, t_{n'-1}$, we have
\begin{enumerate}
\item Fix $\mu$ to arbitrary small value, there has
{\small{\begin{equation}
\lim_{n\rightarrow\infty,n'\rightarrow\infty}\Pr\left\{ \frac{\beta_{1}-\alpha_{1}}{n'-2}\sum_{j=1}^{n'-1}\left|F_{V^{n}\left|X_{1}^{n}\right.}^{n}\left(t_{j}\left|\msf{x}_{1}\right.\right)-\int_{-\infty}^{+\infty}f_{U\left|X_{1}\right.}\left(u\left|\msf{x}_{1}\right.\right)F^{(n')}_{V^{n}\left|\widetilde{U}^{n}\right.}\left(t_{j}\left|u\right.\right)du\right|>\mu\right\} =0.
\end{equation}}}
\item Fix $n'$ to arbitrary large value, and $\epsilon$ to arbitrary small value, there has
{\small{\begin{equation}
\Pr\left\{ \frac{\beta_{1}-\alpha_{1}}{n'-2}\sum_{j=1}^{n'-1}\left|F_{V^{n}\left|X_{1}^{n}\right.}^{n}\left(t_{j}\left|\msf{x}_{1}\right.\right)-\int_{-\infty}^{+\infty}f_{U\left|X_{1}\right.}\left(u\left|\msf{x}_{1}\right.\right)F^{(n')}_{V^{n}\left|\widetilde{U}^{n}\right.}\left(t_{j}\left|u\right.\right)du\right|>\mu_{n'}\right\} \leq\epsilon
\end{equation}}}where $n$ approaches to infinity, $\lim_{n'\rightarrow\infty}\mu_{n'}=0$.
 \end{enumerate}
\end{lemma}
\begin{IEEEproof}
For arbitrary small $\mu$, we have
\begin{align}
&\nonumber \Pr\left\{ \frac{\beta_{1}-\alpha_{1}}{n'-2}\sum_{j=1}^{n'-1}\left|F_{V^{n}\left|X_{1}^{n}\right.}^{n}\left(t_{j}\left|\msf{x}_{1}\right.\right)-\int_{-\infty}^{+\infty}f_{U\left|X_{1}\right.}\left(u\left|\msf{x}_{1}\right.\right)F^{(n')}_{V^{n}\left|\widetilde{U}^{n}\right.}\left(t_{j}\left|u\right.\right)du\right|>\mu\right\}\\
&\nonumber\leq\sum_{j=1}^{n'-1}\Pr\left\{ \left|F_{V^{n}\left|X_{1}^{n}\right.}^{n}\left(t_{j}\left|\msf{x}_{1}\right.\right)-\int_{-\infty}^{+\infty}f_{U\left|X_{1}\right.}\left(u\left|\msf{x}_{1}\right.\right)F^{(n')}_{V^{n}\left|\widetilde{U}^{n}\right.}\left(t_{j}\left|u\right.\right)du\right|>\frac{\mu(n'-2)}{(\beta_{1}-\alpha_{1})(n'-1)}\right\} \\
&\nonumber \leq \underbrace{\frac{4(n'-1)^{2}}{\mu^{2}(n'-2)^{2}}\left(\frac{n'^{3}\left(\beta_{1}-\alpha_{1}\right)^{2}}{n}+\frac{\left(\beta_{1}-\alpha_{1}\right)^{2}n'}{\Pr\left\{ \left(X_{1}^{n},\,\widetilde{U}^{n}\right)\in\mathcal{A}_{\varepsilon}\right\} \left(P_{X_{1}}\left(\mathsf{x}\right)-\varepsilon\right)^{2}}\triangle F_{max}\left(n'\right)\right)+n'\Pr\left\{ \left(X_{1}^{n},\,\widetilde{U}^{n}\right)\notin T_{\left[X_{1},\widetilde{U}\right]_{\varepsilon}}^{n}\right\} }_{\varTheta\left(\mu,n,n'\right)}
\end{align}where the last inequality follows lemma 2. From lemma 1, $\lim_{n'\rightarrow\infty}\left(\beta_{1}-\alpha_{1}\right)^{2}n'\triangle F_{max}\left(n'\right)=0$, then we have 
{\small{\begin{equation}
\lim_{n\rightarrow\infty,n'\rightarrow\infty}\varTheta\left(\mu,n,n'\right)=0
\end{equation}}}for arbitrary value of $\mu$. The first statement of this lemma is proved.
Furthermore, according to the expression of $\varTheta\left(\mu,n,n'\right)$, 
$\varTheta\left(\mu,n,n'\right)<\epsilon$ would be yielded by
\begin{equation}
\frac{4(n'-1)^{2}\left(\beta_{1}-\alpha_{1}\right)^{2}n'\triangle F_{max}\left(n'\right)}{(n'-2)^{2} \left(P_{X_{1}}\left(\mathsf{x}\right)-\varepsilon\right)^{2}\epsilon_{1}}\leq\mu^2
\end{equation}$n\rightarrow\infty$, $\epsilon_{1}<\epsilon$. Hence, by setting 
 \begin{equation}
\frac{4(n'-1)^{2}\left(\beta_{1}-\alpha_{1}\right)^{2}n'\triangle F_{max}\left(n'\right)}{(n'-2)^{2} \left(P_{X_{1}}\left(\mathsf{x}\right)-\varepsilon\right)^{2}\epsilon_{1}}=\mu_{n'}^2
\end{equation}the second statement can be proved.
\end{IEEEproof}
\begin{lemma}\label{dis_nm}
There exist $n_0$, by which for arbitrary $n'>n_0$ we have sequence $\alpha_1=t_{1}<t_{2}<\ldots<t_{n'-1}=\beta_1$ making equation {\small{$$\frac{\beta_1-\alpha_1}{{n'-2}}\sum_{j=1}^{{n'-1}}\left|\int_{-\infty}^{t_{j}}f_{U\left|X_{1}\right.}\left(u\left|\msf{x}_{1}\right.\right)du-\sum_{i}^{n'}P_{\widetilde{U}\left|X_{1}\right.}\left(\widetilde{\mathsf{u}}_{i}\right)w_{i,j}\right|^2=0$$}} have single solution that $w_{i,j}=\Phi\left(t_{j}-\widetilde{\mathsf{u}}_{i}\right),i=1,\ldots n', j=1,\ldots n'-1$ in the domain {\small{$\mathcal{D}=\big\{ w_{i,j}:\,0\leq w_{i,1}\leq w_{i,2}\leq\ldots w_{i,\widehat{n}}\leq1,\, i=1,\ldots n', j=1,\ldots n'-1\big\} $}}.
\end{lemma}
\begin{IEEEproof}
Let us choose $t_{i}=\widetilde{\mathsf{u}}_{j}$, for $j=1,2,\ldots,n'-1$. Then, the proof follows the logic of the proof that wireless channel
is non-manipulable. 
\end{IEEEproof}
\section{Proof of Theorem 1}
Let us go back to the proof of theorem 1. With the aforementioned lemmas, we will show the decision statistic $D^{n}=\frac{1}{{n'-2}}\sum_{j=1}^{{n'-1}}\left|F_{V^{n}\left|X_{1}^{n}\right.}^{n}\left(t_{j}\left|\msf{x}_{1}\right.\right)-\int_{-\infty}^{t_{j}}f_{U\left|X_{1}\right.}\left(u\left|\msf{x}_{1}\right.\right)du\right|$ simultaneously satisfies the properties stated by theorem 1.

From aforementioned work, it is not hard to find
many variables, such as $\alpha_{1}$, $\beta_{1}$ and $\widetilde{\mathsf{u}}$, depend on $n'$.
For easy description,  $n'$ does not appear in these notations. However,
the dependency between $n'$ and these variables will be utilized in the proof given below.
Hence, these notations are written with a superscript $n'$ or possible value of $n'$ so as to
highlight the dependency on $n'$. To be more specific, $n'$ takes value 
from sequence $n'_1, n'_2,\ldots, \infty$, there has $\frac{\triangle_{n'_{k}}}{\triangle_{n'_{k-1}}}=k$ and 
$\frac{\triangle_{n'_{k}}}{\triangle_{n'_{1}}}=s_{k}$, where
$\triangle_{n'_{k}}=\frac{\alpha_{1}^{(n'_{k})}-\beta_{1}^{(n'_{k})}}{n'_{k}-2}$. Upon $n'$, we define
function $M^{\left(n'\right)}\left(W^{\left(n'\right)}\right)=\frac{\beta_{1}^{(n')}-\alpha_{1}^{(n')}}{{n'-2}}\sum_{j=1}^{{n'-1}}\left|\int_{-\infty}^{t_{j}}f_{U\left|X_{1}\right.}\left(u\left|\msf{x}_{1}\right.\right)du-\sum_{i}^{n'}P_{\widetilde{U}\left|X_{1}\right.}\left(\widetilde{\mathsf{u}}_{i}^{(n')}\right)w_{i,j}^{(n')}\right|^{2}$ where $W^{\left(n'\right)}$ is a matrix variable $\left[W^{\left(n'\right)}\right]_{i,j}=w_{i,j}^{(n')}$, $i=1,2,\ldots,n'$, $j=1,2,\ldots,n'-1$. 
As stated by lemma \ref{dis_nm}, $M^{\left(n'\right)}\left(W^{\left(n'\right)}\right)=0$ has single solution
in the point that $W_{0}^{\left(n'\right)}$ defined as $\left[W_{0}^{\left(n'\right)}\right]_{i,j}=\Phi\left(t_{j}^{\left(n'\right)}-\widetilde{\mathsf{u}}_{i}^{(n')}\right)$,  $i=1,\ldots n', j=1,\ldots n'-1$ in the domain $\mathcal{D}^{(n')}=\big\{ W^{(n')}:\,0\leq w^{(n')}_{i,1}\leq w^{(n')}_{i,2}\leq\ldots w^{(n')}_{i,n'-1}\leq1,\, i=1,\ldots n'\big\}$. 
\begin{lemma}\label{lem5}
If $n'$ is sufficient large and $W^{(n')}\in\mathcal{D}_{s}^{(n')}$, where {\small{$\mathcal{D}_{s}^{(n')}=\big\{ W^{(n')}:\,\left|W^{(n')}-W_{0}^{(n')}\right|\geq\delta,\, W^{(n')}\in\mathcal{D}^{(n')}\big\}$}}, then $M^{\left(n'\right)}\left(W^{\left(n'\right)}\right)$ has positive infimum across ${D}_{s}^{(n')}$, denoted as $\lambda^{(n')}\left(\delta\right)$. Moreover, $\lambda^{(n')}\left(\delta\right)\rightarrow0,\delta\rightarrow0$.
\end{lemma}
\begin{IEEEproof}
Using the assertion given by lemma \ref{dis_nm}, the proof of this lemma follows our previous work.
\end{IEEEproof}
\begin{lemma}\label{lem6}
If the wireless channel is non-manipulable, then for arbitrary small $\delta$, there exist sufficient large $n_0$, such that for any $n'>n_{0}$, $\lambda^{(n')}\left(\delta\right)\geq\mu_{n'}$.
\end{lemma}
\begin{IEEEproof}
First notice that for arbitrary $W_{f}^{(n')}\in\mathcal{D}^{(n')}$, 
there exist a CDF function $F\left(t\left|u\right.\right)$ which satisfies
{\small{\begin{equation}
\left[W_{f}^{(n')}\right]_{i,j}=\frac{\int_{u\in\mathcal{B}\left(\widetilde{u}_{i}^{(n')}\right)}F\left(t_{j}^{(n')}\left|u\right.\right)f_{U\left|\mathsf{x}\right.}\left(u\right)du}{\int_{u\in\mathcal{B}\left(\widetilde{u}_{i}^{(n')}\right)}f_{U\left|\mathsf{x}\right.}\left(u\right)du}.
\end{equation}}}Then, according to the condition that 
$\frac{\triangle_{n'_{k}}}{\triangle_{n'_{k-1}}}=k$ and 
$\frac{\triangle_{n'_{k}}}{\triangle_{n'_{1}}}=s_{k}$, fixing $F\left(t\left|u\right.\right)$,
$\left|W_{f}^{(n')}-W_{0}^{(n')}\right|\geq\delta$ implies $\left|W_{f}^{(n'_{k})}-W_{0}^{(n'_{k})}\right|\geq\delta$
for $n'_{k}>n'$.
For the sake of proof, we define a function set 
{\small{\begin{equation}
\mathcal{F}=\left\{ F\left(t\left|u\right.\right):\lim_{n'\rightarrow\infty}\sum_{j=1}^{{n'-1}}\sum_{i=1}^{{n'}}\left|\frac{\int_{u\in\mathcal{B}\left(\widetilde{\mathsf{u}}_{i}^{(n')}\right)}F\left(t_{j}^{(n')}\left|u\right.\right)f_{U\left|\mathsf{x}\right.}\left(u\right)du}{\int_{u\in\mathcal{B}\left(\widetilde{\mathsf{u}}_{i}^{(n')}\right)}f_{U\left|\mathsf{x}\right.}\left(u\right)du}-\Phi\left(t_{j}^{\left(n'\right)}-\widetilde{\mathsf{u}}_{i}^{(n')}\right)\right|\geq\delta\right\}.
\end{equation}}}Then, we define $\mathcal{\widetilde{D}}_{s}^{(n')}$ as
{\small{\begin{equation}
\mathcal{\widetilde{D}}_{s}^{(n')}=\left\{ W^{(n')}:\left[W^{(n')}\right]_{i,j}=\frac{\int_{u\in\mathcal{B}\left(\widetilde{\mathsf{u}}_{i}^{(n')}\right)}F\left(t_{j}^{(n')}\left|u\right.\right)f_{U\left|\mathsf{x}\right.}\left(u\right)du}{\int_{u\in\mathcal{B}\left(\widetilde{\mathsf{u}}_{i}^{(n')}\right)}f_{U\left|\mathsf{x}\right.}\left(u\right)du},\, F\left(t\left|u\right.\right)\in\mathcal{F}\right\} 
\end{equation}}}Obviously, $\mathcal{D}_{s}^{(n')}\subseteq\mathcal{\widetilde{D}}_{s}^{(n')}$, hence, we get $\lambda^{(n')}\left(\delta\right)\geq\widetilde{\lambda}^{(n')}\left(\delta\right)$ where $\widetilde{\lambda}^{(n')}\left(\delta\right)$ is infimum of $M^{\left(n'\right)}\left(W^{\left(n'\right)}\right)$ across $\mathcal{\widetilde{D}}_{s}^{(n')}$. 

In order to prove $\widetilde{\lambda}^{(n')}\left(\delta\right)>\mu_{n'}$, for arbitrary $F\left(t\left|u\right.\right)\in\mathcal{F}$, we assume for arbitrary large $n'_0$, there exist $n'>n'_0$, such that 
$M^{\left(n'\right)}\left(W_{f}^{\left(n'\right)}\right)\leq\mu_{n'}$. In other words, 
there exist a sequence denoted as $\widehat{n}_1<
\widehat{n}_2<\ldots, \infty$ by which 
{\small{\begin{equation}
M^{\left(\widehat{n}_k\right)}\left(W_{f}^{\left(\widehat{n}_k\right)}\right)\leq\mu_{\widehat{n}_k}
\end{equation}}}Then, we have
\begin{equation}
\lim_{k\rightarrow\infty}M^{\left(\widehat{n}_{k}\right)}\left(W_{f}^{\left(\widehat{n}_{k}\right)}\right)\leq\lim_{k\rightarrow\infty}\mu_{\widehat{n}_{k}}=0
\end{equation}From the expressions of $W_{f}^{(n')}$ and $M^{\left(n'\right)}\left(W^{\left(n'\right)}\right)$, we get there is a division manner for $t\in\left(-\infty,+\infty\right)$ characterized by $\widehat{n}_1<
\widehat{n}_2<\ldots, \infty$ such that
{\small{\begin{equation}\label{key1}
\lim_{k\rightarrow\infty}\frac{\beta_{1}^{(\widehat{n}_{k})}-\alpha_{1}^{(\widehat{n}_{k})}}{{\widehat{n}_{k}-2}}\sum_{j=1}^{{\widehat{n}_{k}-2}}\left|\int_{-\infty}^{t_{j}}f_{U\left|X_{1}\right.}\left(u\left|\msf{x}_{1}\right.\right)du-\int_{-\infty}^{t_{j}}f_{U\left|X_{1}\right.}\left(u\right)F\left(t\left|u\right.\right)du\right|^{2}=0
\end{equation}}}
On the other hand, from the definition of $\mathcal{F}$ and the condition that the wireless channel is non-manipulable, we get
\begin{equation}
\int_{-\infty}^{t}f_{U\left|X_{1}\right.}\left(u\left|\msf{x}_{1}\right.\right)du\neq\int_{-\infty}^{t}f_{U\left|X_{1}\right.}\left(u\right)F\left(t\left|u\right.\right)du
\end{equation}Hence, if $\int_{-\infty}^{\infty}\left|\int_{-\infty}^{t}f_{U\left|X_{1}\right.}\left(u\left|\msf{x}_{1}\right.\right)du-\int_{-\infty}^{t}f_{U\left|X_{1}\right.}\left(u\right)F\left(t\left|u\right.\right)du\right|^{2}dt$ can be integrated, we must have
\begin{equation}
\int_{-\infty}^{\infty}\left|\int_{-\infty}^{t}f_{U\left|X_{1}\right.}\left(u\left|\msf{x}_{1}\right.\right)du-\int_{-\infty}^{t}f_{U\left|X_{1}\right.}\left(u\right)F\left(t\left|u\right.\right)du\right|^{2}dt>0,
\end{equation}which indicates there is no division manner for $t\in\left(-\infty,+\infty\right)$ making (\ref{key1}) be true. It contradicts with the meaning
of (\ref{key1}).
We proceed to examine another case that if {\small{$\int_{-\infty}^{\infty}|\int_{-\infty}^{t}f_{U\left|X_{1}\right.}\left(u\left|\msf{x}_{1}\right.\right)du-\int_{-\infty}^{t}f_{U\left|X_{1}\right.}\left(u\right)F\left(t\left|u\right.\right)du|^{2}dt$}} cannot be integrated, since {\small{$\left|\int_{-\infty}^{t}f_{U\left|X_{1}\right.}\left(u\left|\msf{x}_{1}\right.\right)du-\int_{-\infty}^{t}f_{U\left|X_{1}\right.}\left(u\right)F\left(t\left|u\right.\right)du\right|>0$}}, we have
{\small{\begin{equation}
\lim_{\beta\rightarrow\infty,\alpha\rightarrow-\infty}\int_{\alpha}^{\beta}\left|\int_{-\infty}^{t}f_{U\left|X_{1}\right.}\left(u\left|\msf{x}_{1}\right.\right)du-\int_{-\infty}^{t}f_{U\left|X_{1}\right.}\left(u\right)F\left(t\left|u\right.\right)du\right|^{2}dt=\infty
\end{equation}}}Hence, there has $\alpha'$ and $\beta'$ by which 
\begin{equation}\label{key2}
\int_{\alpha'}^{\beta'}\left|\int_{-\infty}^{t}f_{U\left|X_{1}\right.}\left(u\left|\msf{x}_{1}\right.\right)du-\int_{-\infty}^{t}f_{U\left|X_{1}\right.}\left(u\right)F\left(t\left|u\right.\right)du\right|^{2}dt>0
\end{equation} 
Meanwhile, (\ref{key1}) indicates
\begin{equation}\label{key3}
\lim_{k\rightarrow\infty}\frac{\beta_{1}^{(\widehat{n}_{k})}-\alpha_{1}^{(\widehat{n}_{k})}}{{\widehat{n}_{k}-2}}\sum_{t_{j}\in\left[\alpha',\beta'\right]}\left|\int_{-\infty}^{t_{j}}f_{U\left|X_{1}\right.}\left(u\left|\msf{x}_{1}\right.\right)du-\int_{-\infty}^{t_{j}}f_{U\left|X_{1}\right.}\left(u\right)F\left(t\left|u\right.\right)du\right|^{2}=0.
\end{equation}However, (\ref{key2}) indicates there is no division manner for $t\in\left(\alpha',\beta'\right)$ making (\ref{key3}) be true. 
Hence, the contradiction happens. Due to these contradictions, we attain the assumption that for arbitrary large $n'_0$, there exist $n'>n'_0$, such that 
$M^{\left(n'\right)}\left(W_{f}^{\left(n'\right)}\right)\leq\mu_{n'}$ is not right. Therefore, we have there exist $n'_0$, for any $n'>n'_0$, there has $M^{\left(n'\right)}\left(W_{f}^{\left(n'\right)}\right)>\mu_{n'}$. 
Applying the aforementioned derivation to each function belonging 
to $\mathcal{F}$, we get there exist $n_0$, for any $n'>n_0$, $M^{\left(n'\right)}\left(W_{f}^{\left(n'\right)}\right)>\mu_{n'}$ is available 
for all possible functions of $\mathcal{F}$. Since $\widetilde{\lambda}^{(n')}\left(\delta\right)$ is infimum of $M^{\left(n'\right)}\left(W^{\left(n'\right)}\right)$ across $\mathcal{\widetilde{D}}_{s}^{(n')}$, we thus have 
\begin{equation}
\widetilde{\lambda}^{(n')}>\mu_{n'}
\end{equation}Revisiting $\lambda^{(n')}\left(\delta\right)\geq\widetilde{\lambda}^{(n')}\left(\delta\right)$, we get
\begin{equation}
{\lambda}^{(n')}\left(\delta\right)>\mu_{n'}
\end{equation}
Finally, the proof is completed. 
\end{IEEEproof}

\begin{lemma}\label{p2}
Fixing arbitrary small $\epsilon$ and $\delta$, if there exist $n'_0$ such that {\small{$$\Pr\{\underbrace{\sum_{j=1}^{{n'_0-1}}\sum_{i=1}^{n'_0}\left|F_{V^{n}\left|\widetilde{U}^{n}\right.}^{\left(n'_0\right)}\left(t_{j}^{(n'_0)}\left|\widetilde{\mathsf{u}}_{i}^{(n'_0)}\right.\right)-\Phi\left(t_{j}^{(n'_0)}-\widetilde{\mathsf{u}}_{i}^{(n'_0)}\right)\right|}_{R\left(U^{n},V^{n},n'_0\right)}>\delta\} >0$$}}then, we have for $n'>n'_0$, {\small{$\Pr\left\{ D^{n}<\varepsilon\left(n', \delta\right)\left|{R\left(U^{n},V^{n},n'\right)}>\delta\right.\right\}$}} is well-defined and 
$$\Pr\left\{ D^{n}<\varepsilon\left(n', \delta\right)\left|{R\left(U^{n},V^{n},n'\right)}>\delta\right.\right\} \leq\epsilon$$ where $n\rightarrow\infty$, 
$n'$ is sufficient large so as to satisfy the properties given by lemma 4 and lemma 6. $\varepsilon\left(n', \delta\right)$ is strictly positive and can be arbitrary small value.
\end{lemma}

\begin{IEEEproof}
According to lemma 3, there exist $\mu_{n'}$ such that
{\small{\begin{align}\label{p2_1}
\nonumber&\Pr\left\{ \frac{\beta^{(n')}_1-\alpha^{(n')}_1}{{n'-2}}\sum_{j=1}^{{n'-1}}\left|F_{V^{n}\left|X_{1}^{n}\right.}^{n}\left(t^{(n')}_{j}\left|\msf{x}_{1}\right.\right)-\sum_{i}^{n'}P_{\widetilde{U}\left|X_{1}\right.}\left(\widetilde{\mathsf{u}}^{(n')}_{i}\right)F_{V^{n}\left|\widetilde{U}^{n}\right.}^{\left(n'\right)}\left(t^{(n')}_{j}\left|\widetilde{\mathsf{u}}^{(n')}_{i}\right.\right)\right|>\mu_{n'}\right\}\\& \leq\epsilon\Pr\left\{ R\left(U^{n},V^{n},n'_{0}\right)>\delta\right\}\leq\epsilon\Pr\left\{ R\left(U^{n},V^{n},n'\right)>\delta\right\},
\end{align}}}where $\mu_{n'}\rightarrow0$ as $n'\rightarrow\infty, n\rightarrow\infty$. The last inequality follows the fact that $R\left(U^{n},V^{n},n'_{0}\right)>\delta$ implies $R\left(U^{n},V^{n},n'\right)>\delta$ according to the definition of $F_{V^{n}\left|\widetilde{U}^{n}\right.}^{\left(n'\right)}\left(t^{(n')}_{j}\left|\widetilde{\mathsf{u}}^{(n')}_{i}\right.\right)$. 

Then, if {\small{$$\underbrace{\frac{\beta_{1}^{(n')}-\alpha_{1}^{(n')}}{{n'-2}}\sum_{j=1}^{{n'-1}}\left|F_{V^{n}\left|X_{1}^{n}\right.}^{n}\left(t_{j}^{(n')}\left|\msf{x}_{1}\right.\right)-\sum_{i}^{n'}P_{\widetilde{U}\left|X_{1}\right.}\left(\widetilde{\mathsf{u}}_{i}^{(n')}\right)F_{V^{n}\left|\widetilde{U}^{n}\right.}^{\left(n'\right)}\left(t_{j}^{(n')}\left|\widetilde{\mathsf{u}}_{i}^{(n')}\right.\right)\right|}_{G\left(V^{n},U^{n},X_{1}^{n}\right)}<\mu_{n'},$$}}we must have
{\small{\begin{align}\label{p2_2}
&\left(\beta^{(n')}_1-\alpha^{(n')}_1\right)D^{n}\geq\frac{\beta^{(n')}_1-\alpha^{(n')}_1}{n'-2}\sum_{j=1}^{{n'-1}}\left|\int_{-\infty}^{t^{(n')}_{j}}f_{U\left|X_{1}\right.}\left(u\left|\msf{x}_{1}\right.\right)du-\sum_{i}^{n'}P_{\widetilde{U}\left|X_{1}\right.}\left(\widetilde{\mathsf{u}}^{(n')}_{i}\right)F_{V^{n}\left|\widetilde{U}^{n}\right.}^{\left(n'\right)}\left(t^{(n')}_{j}\left|\widetilde{\mathsf{u}}^{(n')}_{i}\right.\right)\right|-\mu_{n'}\\\nonumber
&\geq\frac{\beta^{(n')}_1-\alpha^{(n')}_1}{n'-2}\sum_{j=1}^{{n'-1}}\left|\int_{-\infty}^{t^{(n')}_{j}}f_{U\left|X_{1}\right.}\left(u\left|\msf{x}_{1}\right.\right)du-\sum_{i}^{n'}P_{\widetilde{U}\left|X_{1}\right.}\left(\widetilde{\mathsf{u}}^{(n')}_{i}\right)F_{V^{n}\left|\widetilde{U}^{n}\right.}^{\left(n'\right)}\left(t^{(n')}_{j}\left|\widetilde{\mathsf{u}}^{(n')}_{i}\right.\right)\right|^2-\mu_{n'}
\end{align}}}On the other hand, if {\small{$\underbrace{\sum_{j=1}^{{n'-1}}\sum_{i=1}^{n'}\left|F_{V^{n}\left|\widetilde{U}^{n}\right.}^{\left(n'\right)}\left(t_{j}^{(n')}\left|\widetilde{\mathsf{u}}_{i}^{(n')}\right.\right)-\Phi\left(t_{j}^{(n')}-\widetilde{\mathsf{u}}_{i}^{(n')}\right)\right|}_{R\left(U^{n},V^{n},n'\right)}>\delta,$}}   according to lemma \ref{lem5}, the right side of (\ref{p2_2}) becomes
{\small{\begin{equation}\label{p2_3}
\left(\beta_{1}^{(n')}-\alpha_{1}^{(n')}\right)D^{n}\geq\lambda^{(n')}\left(\delta\right)-\mu_{n'}
\end{equation}}}which can be reshaped as 
{\small{\begin{equation}\label{p2_4}
D^{n}\geq\frac{\lambda^{(n')}\left(\delta\right)-\mu_{n'}}{\left(\beta_{1}^{(n')}-\alpha_{1}^{(n')}\right)}
\end{equation}}}Define $\varepsilon\left(n',\delta\right)=\frac{\lambda^{(n')}\left(\delta\right)-\mu_{n'}}{\left(\beta_{1}^{(n')}-\alpha_{1}^{(n')}\right)}$,
according to lemma \ref{lem6}, $\varepsilon\left(n',\delta\right)>0$ as $n'$ is sufficient large.
From the properties of $\mu_{n'}$ and $\lambda^{(n')}\left(\delta\right)$, 
$\varepsilon\left(n',\delta\right)$ can be arbitrarily small.  

Upon (\ref{p2_2}), (\ref{p2_3}) and (\ref{p2_4}), we have
{\small{\begin{align*}
&\Pr\left\{ D^{n}\geq\varepsilon\left(n',\delta\right),R\left(U^{n},V^{n},n'\right)\geq\delta\right\} \\
&\geq\Pr\left\{ D^{n}\geq\varepsilon\left(n',\delta\right),R\left(U^{n},V^{n},n'\right)\geq\delta,G\left(V^{n},U^{n},X_{1}^{n}\right)\leq\mu'\right\}\\
&=\Pr\left\{ R\left(U^{n},V^{n},n'\right)\geq\delta,G\left(V^{n},U^{n},X_{1}^{n}\right)\leq\mu_{n'}\right\}\\ 
&\geq\Pr\left\{ R\left(U^{n},V^{n},n'\right)\geq\delta\right\} -\Pr\left\{ G\left(V^{n},U^{n},X_{1}^{n}\right)\geq\mu_{n'}\right\} 
\end{align*}}}where the equation follows the logic from (\ref{p2_2}), (\ref{p2_3}) to (\ref{p2_4}). Then, we have
{\small{\begin{equation}
\Pr\left\{ D^{n}\geq\varepsilon\left(n',\delta\right)\left|R\left(U^{n},V^{n},n'\right)\geq\delta\right.\right\} =\frac{\Pr\left\{ D^{n}\geq\varepsilon\left(n',\delta\right),R\left(U^{n},V^{n},n'\right)\geq\delta\right\} }{\Pr\left\{ R\left(U^{n},V^{n},n'\right)\geq\delta\right\} }>1-\frac{\Pr\left\{ G\left(V^{n},U^{n},X_{1}^{n}\right)\geq\mu_{n'}\right\} }{\Pr\left\{ R\left(U^{n},V^{n},n'\right)\geq\delta\right\} }\geq1-\epsilon
\end{equation}}}where the last inequality follows (\ref{p2_1}).
The proof is finished.
\end{IEEEproof}
The first property of theorem 1 is direct result from lemma {\ref{p2}}. 


We proceed to prove the second property of theorem 1. For arbitrary small $\delta$, $\mu$ and $\mu'(n', \delta)=\mu+\frac{\delta}{{n'-2}}$, we have
{\small{\begin{align}
&\nonumber \Pr\left\{ D^{n}\leq\mu'(n', \delta)\bigcap\sum_{j=1}^{{n'-1}}\sum_{i=1}^{n'}\left|F^{(n')}_{V^{n}\left|\widetilde{U}^{n}\right.}\left(t^{(n')}_{j}\left|\widetilde{\mathsf{u}}^{(n')}_{i}\right.\right)-\Phi\left(t^{(n')}_{j}-\widetilde{\mathsf{u}}^{(n')}_{i}\right)\right|\leq\delta\right\} \geq\\
&\nonumber \Pr\big\{ D^{n}\leq\mu'(n', \delta)\bigcap\frac{1}{{n'-2}}\sum_{j=1}^{{n'-1}}\left|F_{V^{n}\left|X_{1}^{n}\right.}^{n}\left(t^{(n')}_{j}\left|\msf{x}_{1}\right.\right)-\int_{-\infty}^{+\infty}f_{U\left|X_{1}\right.}\left(u\left|\msf{x}_{1}\right.\right)F^{(n')}_{V^{n}\left|\widetilde{U}^{n}\right.}\left(t^{(n')}_{j}\left|u\right.\right)du\right|\leq\mu\\&\nonumber \bigcap\sum_{j=1}^{{n'-1}}\sum_{i=1}^{n'}\left|F^{(n')}_{V^{n}\left|\widetilde{U}^{n}\right.}\left(t^{(n')}_{j}\left|\widetilde{\mathsf{u}}^{(n')}_{i}\right.\right)-\Phi\left(t^{(n')}_{j}-\widetilde{\mathsf{u}}^{(n')}_{i}\right)\right|\leq\delta\big\} \\
&\nonumber =\Pr\big\{ \frac{1}{{n'-2}}\sum_{j=1}^{{n'-1}}\left|F_{V^{n}\left|X_{1}^{n}\right.}^{n}\left(t^{(n')}_{j}\left|\msf{x}_{1}\right.\right)-\int_{-\infty}^{+\infty}f_{U\left|X_{1}\right.}\left(u\left|\msf{x}_{1}\right.\right)F^{(n')}_{V^{n}\left|\widetilde{U}^{n}\right.}\left(t^{(n')}_{j}\left|u\right.\right)du\right|\leq\mu\\&\nonumber \bigcap\sum_{j=1}^{{n'-1}}\sum_{i=1}^{n'}\left|F^{(n')}_{V^{n}\left|\widetilde{U}^{n}\right.}\left(t^{(n')}_{j}\left|\widetilde{\mathsf{u}}^{(n')}_{i}\right.\right)-\Phi\left(t^{(n')}_{j}-\widetilde{\mathsf{u}}^{(n')}_{i}\right)\right|\leq\delta\big\} \\
&\nonumber \geq\Pr\left\{\sum_{j=1}^{{n'-1}}\sum_{i=1}^{n'}\left|F^{(n')}_{V^{n}\left|\widetilde{U}^{n}\right.}\left(t^{(n')}_{j}\left|\widetilde{\mathsf{u}}^{(n')}_{i}\right.\right)-\Phi\left(t^{(n')}_{j}-\widetilde{\mathsf{u}}^{(n')}_{i}\right)\right|\leq\delta\right\} -\\&\nonumber\Pr\left\{  \frac{1}{{n'-2}}\sum_{j=1}^{{n'-1}}\left|F_{V^{n}\left|X_{1}^{n}\right.}^{n}\left(t^{(n')}_{j}\left|\msf{x}_{1}\right.\right)-\int_{-\infty}^{+\infty}f_{U\left|X_{1}\right.}\left(u\left|\msf{x}_{1}\right.\right)F^{(n')}_{V^{n}\left|\widetilde{U}^{n}\right.}\left(t^{(n')}_{j}\left|u\right.\right)du\right|>\mu\right\}\label{p1}
\end{align}}}where the equality firstly follows the fact that according to $t^{(n')}_j \in\widetilde{\mathcal{U}}$ and the definition of {\small{$F^{(n')}_{V^{n}\left|\widetilde{U}^{n}\right.}\left(t\left|u\right.\right)$}}, {\small{$$\sum_{j=1}^{{n'-1}}\sum_{i=1}^{n'}\left|F^{(n')}_{V^{n}\left|\widetilde{U}^{n}\right.}\left(t^{(n')}_{j}\left|\widetilde{\mathsf{u}}^{(n')}_{i}\right.\right)-\Phi\left(t^{(n')}_{j}-\widetilde{\mathsf{u}}^{(n')}_{i}\right)\right|\leq\delta$$}} indicates {\small{$\sum_{j=1}^{{n'-1}}\sup_{u}\left|F_{V^{n}\left|\widetilde{U}^{n}\right.}\left(t^{(n')}_{j}\left|u\right.\right)-\Phi\left(t^{(n')}_{j}-u\right)\right|\leq\delta$,}} combining with another event $$\frac{1}{{n'-2}}\sum_{j=1}^{{n'-1}}\left|F_{V^{n}\left|X_{1}^{n}\right.}^{n}\left(t^{(n')}_{j}\left|\msf{x}_{1}\right.\right)-\int_{-\infty}^{+\infty}f_{U\left|X_{1}\right.}\left(u\left|\msf{x}_{1}\right.\right)F^{(n')}_{V^{n}\left|\widetilde{U}^{n}\right.}\left(t^{(n')}_{j}\left|u\right.\right)du\right|\leq\mu,$$ we have
\begin{align}
&\nonumber D^{n}\leq\frac{1}{{n'-2}}\sum_{j=1}^{{n'-1}}\left|F_{V^{n}\left|X_{1}^{n}\right.}^{n}\left(t^{(n')}_{j}\left|\msf{x}_{1}\right.\right)-\int_{-\infty}^{+\infty}f_{U\left|X_{1}\right.}\left(u\left|\msf{x}_{1}\right.\right)F^{(n')}_{V^{n}\left|\widetilde{U}^{n}\right.}\left(t^{(n')}_{j}\left|u\right.\right)du\right|\\
&\nonumber+\frac{1}{{n'-2}}\sum_{j=1}^{{n'-1}}\left|\int_{-\infty}^{+\infty}f_{U\left|X_{1}\right.}\left(u\left|\msf{x}_{1}\right.\right)\left(F^{(n')}_{V^{n}\left|\widetilde{U}^{n}\right.}\left(t_{j}\left|u\right.\right)-\Phi\left(t^{(n')}_{j}-u\right)\right)du\right|\\
&<\mu+\frac{\delta}{{n'-2}}=\mu'(n', \delta).
\end{align}Hence, the equality in (\ref{p1}) is established. Upon (\ref{p1}) and lemma 3, the property 2 in theorem 1 is direct.

\bibliographystyle{IEEEtran}

\end{document}